\documentclass[aps,showpacs,superscriptaddress,preprint,tightenlines]{revtex4}
\usepackage{hyperref}
\usepackage{times}
\usepackage{amsmath}
\usepackage{graphicx}
\usepackage{epsfig}
\usepackage{dcolumn}
\usepackage{bm}
\usepackage{color}
\usepackage{longtable}
\usepackage{array}
\usepackage{multirow}
\newcommand{\PreserveBackslash}[1]{\let\temp=\\#1\let\\=\temp}
\newcolumntype{C}[1]{>{\PreserveBackslash\centering}p{#1}}
\newcolumntype{R}[1]{>{\PreserveBackslash\raggedleft}p{#1}}
\newcolumntype{L}[1]{>{\PreserveBackslash\raggedright}p{#1}}

\newcommand{\Rb}{R_b}
\newcommand{\RbB}{R_b^{\rm B}}
\newcommand{\pp}{\pi^+\pi^-}

\newcommand{\EE}{e^+e^-}
\newcommand{\MM}{\mu^+\mu^-}

\newcommand{\psip}{\psi(3686)}

\newcommand{\jpsi}{J/\psi}

\newcommand{\bbb}{b\bar{b}}

\newcommand{\beq}{\begin{equation}}
\newcommand{\eeq}{\end{equation}}

\newcommand\Eq[1]{Eq.~\eref{#1}}
\def\eref#1{(\ref{#1})}
\newcommand{\rhos}{(1+\delta(s))}

\begin{document}

\title{\boldmath Hadronic cross section of $\EE$ annihilation at bottomonium energy region}

 \affiliation{Institute of High Energy Physics, Chinese Academy of Sciences, Beijing 100049, China}
  \affiliation{Institute of Theoretical Physics, Chinese Academy of Sciences, Beijing 100190, China}
 \affiliation{University of Chinese Academy of Sciences, Beijing 100049, China}
\author{Xiang-Kun Dong}
 \email{dongxiangkun14@mails.ucas.edu.cn}
  \affiliation{Institute of Theoretical Physics, Chinese Academy of Sciences, Beijing 100190, China}
 \affiliation{University of Chinese Academy of Sciences, Beijing 100049, China}
\author{Xiao-Hu Mo}
 \email{moxh@ihep.ac.cn}
 \affiliation{Institute of High Energy Physics, Chinese Academy of Sciences, Beijing 100049, China}
 \affiliation{University of Chinese Academy of Sciences, Beijing 100049, China}
\author{Ping Wang}
 \email{wangp@ihep.ac.cn}
 \affiliation{Institute of High Energy Physics, Chinese Academy of Sciences, Beijing 100049, China}
\author{Chang-Zheng Yuan}
 \email{yuancz@ihep.ac.cn}
 \affiliation{Institute of High Energy Physics, Chinese Academy of Sciences, Beijing 100049, China}
 \affiliation{University of Chinese Academy of Sciences, Beijing 100049, China}

\date{\today}

\begin{abstract}

The Born cross section and dressed cross section of $\EE\to \bbb$ and the total hadronic
cross section in $\EE$ annihilation in the bottomonium energy
region are calculated based on the $\Rb$ values measured by the
BaBar and Belle experiments. The data are used to calculate the
vacuum polarization factors in the bottomonium energy
region, and to determine the resonant parameters of the vector 
bottomonium(-like) states, $Y(10750)$, $\Upsilon(5S)$, and $\Upsilon(6S)$.

\end{abstract}


\maketitle

\section{Introduction}

The cross section of $\EE$ annihilation into hadrons is essential
information for Quantum Electrodynamics (QED) as it is related to the vacuum polarization (VP) of
the photon propagator. The measurement of these cross sections is
one of the important topics in various $\EE$ colliders from low
to high energy, and the precision of the measurements has been
successively improved since the running of the first
generation of $\EE$ colliders~\cite{experiments}. The data have been
used in many calculations involving the photon propagator,
especially in the high precision calculations of the anomalous
magnetic moment of the $\mu$, $a_\mu$, and the running of the fine
structure function, $\alpha(s)$, where $s$ is the center-of-mass 
(CM) energy squared~\cite{Davier,Teubner,ref-ask-3-fred}.

The cross section of $\EE$ annihilation into hadrons is often
reported in terms of $R$ value, defined as
\begin{equation}\label{define_r}
 R=\frac{\sigma^B(\EE\to {\rm hadrons})}{\sigma^B(\EE\to \MM)},
\end{equation}
where $\sigma^B(\EE\to \MM)=\frac{4\pi\alpha^2(0)}{3s}$, is the
Born cross section of $\EE\to \MM$. The experimental measurements
of the $R$ values are compiled in Ref.~\cite{experiments}. There
are many data at low energies ($\sqrt{s}<2$~GeV) with precision at
1\% level; while the measurements are sparse and less precise at
higher energies, for example, the charmonium
($3.7<\sqrt{s}<5.0$~GeV) and the bottomonium
($10.5<\sqrt{s}<11.2$~GeV) energy regions. One of the reasons of
less measurements at high energy is the smaller contribution to
the VP, and another reason is the fact that
fewer experiments were designed in these energy regions.

The cross sections of $\EE\to \bbb$ were measured in
much higher precision by BaBar~\cite{babar_rb} and
Belle~\cite{belle_rb} experiments in the bottomonium energy
region, i.e., $\sqrt{s}=10.5$ to $11.2$~GeV, than by
CUSB~\cite{CUSB} and CLEO~\cite{CLEO} experiments more than 30
years ago. However, neither BaBar nor Belle (let alone CUSB and
CLEO) did radiative corrections to the measured cross sections, so
the data cannot be used directly for many calculations where the
Born cross sections are needed as input.

In this paper, we describe how to get the Born cross section based
on the published data from the BaBar and Belle experiments with some
reasonable assumptions. We report the Born cross sections from
these experiments and discuss the usage of the data samples in 
the calculation of the
VP factors especially in the bottomonium energy
region, and the fit to the dressed cross sections to extract the 
resonant parameters of the vector bottomonium states.
We also discuss a possible determination of the VP directly 
by measuring $\EE\to \MM$ cross sections
with high luminosity data at the Belle or Belle II experiment, and
a strategy to search for the production of invisible particles in
$\EE$ annihilation.

\section{Radiative correction}

The experimentally observed cross section ($\sigma^{\rm obs}$) is
related to the Born cross section via
\begin{equation}
\sigma^{\rm obs} (s)=\int \limits_{0}^{x_m} F(x,s)
\frac{\sigma^{\rm B}(s(1-x))}{|1-\Pi (s(1-x))|^2} \ \mathrm{d}x ,
\label{radsec}
\end{equation}
where $\sigma^{\rm B}$ is the Born cross section, $F(x,s)$ has
been calculated in Refs.~\cite{rad,Berends,ref-ask-2-mont} and $\frac{1}{|1-\Pi(s)|^2}$ 
is the VP factor; the upper limit of the integration
$x_m=1-s_m/s$, where $\sqrt{s_m}$ is the experimentally required
minimum invariant mass of the final state $f$ after losing energy
to multi-photon emission. In this paper, $\sqrt{s_m}$ corresponds
to the $B\bar{B}$ mass threshold, which is $10.5585$~GeV.

The radiator $F(x,s)$ is usually expressed as~\cite{rad}
\begin{align}
F(x,s) =&\ x^{\beta-1}\beta \cdot ( 1+\delta^\prime )
-\beta(1-\frac{1}{2}x)\notag\\
&+\frac{1}{8}\beta^2
\left[4(2-x)\ln\frac{1}{x}-\frac{(1+3(1-x)^2)}{x}\ln(1-x)-6+x\right],
\label{FKuraev}
\end{align}
with
\begin{equation}
\delta^\prime=\frac{\alpha}{\pi} (\frac{\pi^2}{3} - \frac{1}{2})+
\frac{3}{4} \beta + \beta^2(\frac{9}{32}-\frac{\pi^2}{12}),
\label{defdelta}
\end{equation}
and
\begin{equation}
\beta = \frac{2 \alpha}{\pi} \left( \ln \frac{s}{m^2_e} -1
\right). \label{defoft}
\end{equation}
Here the conversion of soft photons into real $\EE$ pairs is
included.

The Born cross section is thus calculated from
\begin{equation}\label{formula-cs}
\sigma^{\rm B}(s)=\frac{\sigma^{\rm obs}(s)}
 {\rhos \cdot \frac{1}{|1-\Pi(s)|^{2}}},
\end{equation}
where $\rhos$ is the initial state radiation (ISR) correction
factor.

It is obvious that both $\rhos$ and $\frac{1}{|1-\Pi(s)|^{2}}$ depend on the
Born cross section from threshold up to the CM energy under study,
while the Born cross section is the quantity we want to measure. These two
factors can only be obtained using the measured quantities with an
iteration procedure.

The pure ISR correction factor $\rhos$ depends only on the line
shape of $\EE\to \bbb$ cross section, while $\frac{1}{|1-\Pi(s)|^{2}}$
depends also on the $R$ values in the full energy range, we use a
two-step procedure to get the Born cross sections.

\subsection{ISR correction factor}

The ISR correction factor is obtained with an iterative procedure, 
following Ref.~\cite{Dong:2017tpt}, via
\begin{align}
  \sigma^{\mathrm{obs}}_{i+1}(s)&=\int_0^{x_{\mathrm{m}}}F(x,s)\sigma^{\mathrm{dre}}_i(s(1-x))\  \mathrm{d}x,\label{isr-factor}\\
  \frac1{1+\delta_{i+1}(s)}&=\sigma^{\mathrm{dre}}_i(s)/\sigma^{\mathrm{obs}}_{i+1}(s),\label{isr-factor2}\\
 \sigma^{\mathrm{dre}}_{i+1}(s)& =
 \frac1{1+\delta_{i+1}(s)} \sigma^{\mathrm{obs}}(s)\label{isr-factorf}
\end{align}
where $\sigma^{\mathrm{dre}}(s)=\frac{\sigma^{\mathrm{B}}(s)}{|1-\Pi(s)|^2}$ is 
the dressed cross section. At the zeroth step of the iteration, the observed 
cross sections are inserted into the integral, playing the role of the dressed 
cross sections, i.e. $\sigma^{\mathrm{dre}}_0(s)=\sigma^{\mathrm{obs}}(s)$. 
The iteration is continued until the
difference between the two consecutive results is smaller than a
given upper limit. The result from the last iteration, denoted by
$(1+\delta_f(s))$, is regarded as the final ISR correction factor.

\subsection{Vacuum polarization factor}

A similar procedure is used to calculate the VP
factor in the bottomonium energy region. In this calculation,
however, the total hadronic cross section is used rather than 
that of $\EE\to \bbb$ only. Moreover, instead of depending
on the hadronic cross sections in the bottomonium energy region,
the VP factor depends on the $R$ values in the
full energy region. In addition, there is also contribution from
leptons.
The VP factor includes two
terms~\cite{Greiner:1994}
 \beq
 \Pi(s) \equiv \sum_{j=e,\,\mu,\,\tau} \Pi_l(s,m^2_j) + \Pi_h(s) ~.
 \label{defdeltav}
 \eeq
The first term is the contribution from the leptonic loops
with
 \beq
 \Pi_l(s,m^2) = \Pi_R+i~\Pi_I
 \label{defpilepton}
 \eeq
for lepton with mass $m$. For $0 \le s <4 m^2$, we define
$a=(4m^2/s-1)^{1/2}$,
 \beq
 \begin{array}{rcl}
 \Pi_R&=&-{\displaystyle \frac{\alpha}{\pi}
 \left[\frac{8}{9}+\frac{a^2}{3}
 -2 \left(\frac{1}{2}+\frac{a^2}{6}\right) \cdot a
 \cdot \cot^{-1}(a)~\right],
 } \\
 \Pi_I&=&{\displaystyle 0~, }
 \end{array}
 \eeq
while for $s \ge 4 m^2 $, we define $a=(1-4m^2/s)^{1/2}$ and
$b=(1-a)/(1+a)$,
 \beq
 \begin{array}{rcl}
 \Pi_R&=&-{\displaystyle \frac{\alpha}{\pi}
 \left[\frac{8}{9}-\frac{a^2}{3} +
 \left(\frac{1}{2}-\frac{a^2}{6}\right) \cdot a
 \cdot \ln b ~ \right], }\\
 \Pi_I&=& -{\displaystyle \frac{a \alpha}{3}
 \left(1+\frac{2 m^2}{s}\right)~. }
 \end{array}
 \eeq

The second term in Eq.~\eref{defdeltav} is the contribution
from the hadronic loops. This quantity $\Pi_h(s)$ is related to
the total cross section $\sigma (s)$ of $\EE \to {\rm hadrons}$
in the one-photon exchange approximation through a dispersion
relation
 \beq
\Pi_h(s)= \frac{s}{4\pi^2 \alpha} \int_{4m^2_{\pi}}^{\infty}
\frac{\sigma (s^{\prime})}{s-s^{\prime}+i\epsilon } d s^{\prime}~.
\label{dspnrelationa}
 \eeq
Using the identity
$$\frac{1}{x+i\epsilon} = P\frac{1}{x}-i\pi \delta(x)~, $$
we have
 \beq
\Pi_h(s)=- \frac{s}{4\pi^2 \alpha} P \int_{4m^2_{\pi}}^{\infty}
\frac{\sigma (s^{\prime})}{s^{\prime}-s} d s^{\prime} - i
\frac{s}{4\pi \alpha} \sigma(s)~. \label{dspnrelationb}
 \eeq

We follow the procedure in Ref.~\cite{Berends_vp} to calculate 
the first term in the above equation. First, the integration 
is performed analytically for
narrow resonances $\jpsi$, $\psip$, $\Upsilon(1S)$,
$\Upsilon(2S)$, and $\Upsilon(3S)$. Second, for the high energy
part, it is assumed that $R(s)=R(s_1)$ is a constant above a
certain value $s_1$. And third, the integral between threshold and
$s_1$ is carried out numerically after separation of the principle
value part. Thus we have
\begin{align}
\Re~ \Pi_h(s)=& \frac{3s}{\alpha} \sum\limits_{j}
\frac{\Gamma^j_{\EE}}{M_j}\frac{s -M_j^2}{(s -M_j^2)^2 + M_j^2 \Gamma_j^2}
+\frac{\alpha}{3\pi} R(s_1)\ln\left|\frac{s-s_1}{s_1}\right|\notag\\
&- \frac{s}{4\pi^2 \alpha}
 \int_{4m^2_{\pi}}^{s_1} \frac{\sigma_{\rm nr} (s^{\prime})-\sigma_{\rm nr} (s)}
 {s^{\prime}-s}\mathrm{d} s^{\prime} -\frac{s\sigma_{\rm nr} (s)}{4\pi^2 \alpha}
 \ln \left|\frac{s_1-s}{4m^2_{\pi}-s}\right| ,\label{eq:piexpressiona}
\end{align}
where $\Gamma_j$, $\Gamma^j_{\EE}$, and $M_j$ denote total width,
partial width to $\EE$ pair, and mass of the resonance $j$,
respectively. Here, $\sigma_{\rm nr} (s)$ is the $\sigma (s)$ 
in~\Eq{dspnrelationb} with the contributions from narrow
resonances subtracted.

We use experimental measurements or theoretical calculations of
$R$ values in different energy regions in the calculation of the
VP factors:
\begin{enumerate}

\item For $2m_{\pi}<\sqrt{s}<0.36$~GeV, we consider $\EE\to \pp$
only, with the $\pi$ form factor obtained
through~\cite{Davier:2002dy}
 \beq
F_{\pi}(s) = 1+ \frac{1}{6} \langle r^2 \rangle_{\pi} s + c_1 s^2
+ c_2 s^3 ~,
 \eeq
where $\langle r^2 \rangle_{\pi} = 0.429$, $c_1 = 6.8$, and
$c_2=-0.7$.

\item For $0.36<\sqrt{s}<2.0$~GeV, we used $R$ values from PDG
compilation~\cite{PDG,experiments}.

\item For $3.7<\sqrt{s}<5.0$~GeV, we use $R$ values from the BES
collaboration~\cite{Bai:1999pk,Bai:2001ct}.

\item For $10.5585<\sqrt{s}<11.2062$~GeV, we use the $\Rb$ values
provided by the Belle and BaBar
collaborations~\cite{babar_rb,belle_rb} with proper handling of
the ISR correction and VP correction described
below.

\item For all the other energy regions, we use $R$ values from
pQCD calculation~\cite{PDG,Rodrigo:1997zd}
 \beq
R_{\rm QCD} (s)=R_{\rm EW}(s) [1+\delta_{\rm QCD}(s)]~, \label{eq:rqcd}
 \eeq
where $R_{\rm EW}(s)=3\Sigma_q e_q^2$ is the purely electroweak
contribution neglecting finite-quark-mass corrections with $e_q$
the electric charges of the quarks; the QCD correction factor is given by
 \beq
\delta_{\rm QCD}(s)= \sum_{i=1}^{4} c_i \left[ \frac{\alpha_s(s)}{\pi}
\right]^i~, \label{eq:deltaqcd}
 \eeq
with parameters defined in Refs.~\cite{PDG,Rodrigo:1997zd}.

\end{enumerate}

Replacing pQCD calculations with recent KEDR measurements~\cite{ref-KEDR1,ref-KEDR2} 
for $\sqrt{s}$ between 2 and 3.7~GeV gives very similar results in the bottomonium 
energy region of interest. 

In the bottomonium energy region, the dressed cross section 
of $\EE\to \bbb$ is denoted by 
$\sigma^{\rm dre}_b(s)=\frac{\sigma^{\rm B}_b(s)} {|1-\Pi(s)|^2}=(1+\delta_f(s))\sigma^{\rm{obs}}(\EE\to \bbb)$ 
where $\sigma^{\rm{obs}}(\EE\to \bbb)$ is the observed cross section provided by the Belle and BaBar
collaborations~\cite{babar_rb,belle_rb}, and the Born cross section of $\EE\to u,d,s,c$-quarks 
from the pQCD calculation is denoted by $\sigma_{udsc}^{\rm B}(s)$. 
Then $\sigma_{0}^{\rm B}(s)=\sigma_{udsc}^{\rm B}(s)+\sigma^{\rm dre}_b(s)$ is taken
as zeroth order approximation of the Born cross section of $\EE\to {\rm hadrons}$. 
Together with the Born cross sections in other energy regions we obtain the first order 
approximation of the VP factor,
$\frac{1}{|1-\Pi_1(s)|^2}$, via Eqs.~(\ref{dspnrelationb}) and
(\ref{eq:piexpressiona}). Then we use 
$\sigma_{i}^{\rm B}(s)=\sigma_{udsc}^{\rm B}(s)+\sigma^{\rm dre}_b(s)/\frac{1}{|1-\Pi_i(s)|^2}$,  
the $i$th order approximation of
$\sigma^{\rm B}(s)$, to calculate $\frac{1}{|1-\Pi_{(i+1)}(s)|^2}$. 
We iterate this procedure until
$\frac{1}{|1-\Pi_i(s)|^2}$ is stable and take it as the final 
VP factor $\frac{1}{|1-\Pi_f(s)|^2}$.

\subsection{Born cross section}

The final Born cross section of $\EE\to \bbb$ can then be
calculated with Eq.~\eref{formula-cs} with the ISR correction
factor and VP factor calculated above, i.e.,
\begin{equation}\label{formula-csf}
\sigma_b^{\rm B}(s)=\frac{\sigma^{\rm obs}(s)}
 {(1+\delta_f(s))\frac{1}{|1-\Pi(s)|^{2}_f}}.
\end{equation}

\section{The data}

Both BaBar~\cite{babar_rb} and Belle~\cite{belle_rb} experiments
measured $\Rb$ in the bottomonium energy region: 
$$R_b\equiv \frac{\sigma(\EE\to b\bar{b})}{\sigma^B(\EE\to \MM)},$$
where the denominator is the Born cross section of $\EE\to \MM$.
In both experiments, neither ISR correction,
nor the VP correction was considered, so the reported
$\Rb$ corresponds to the observed cross section. In both
experiments, the contribution of the narrow $\Upsilon$ states
from the initial state radiation, i.e., $\Upsilon(1S)$, 
$\Upsilon(2S)$, and $\Upsilon(3S)$ states can be removed 
from the data supplied in the papers.

The BaBar measurement~\cite{babar_rb} was based on data collected
between March 28 and April 7, 2008 at CM energies from
10.54 to 11.20~GeV. First, an energy scan over the whole range in
5~MeV steps, collecting approximately 25~pb$^{-1}$ per step for a
total of about 3.3~fb$^{-1}$, was performed. This was then followed
by a 600~pb$^{-1}$ scan in the range of CM energy from 10.96 to
11.10~GeV, in 8 steps with non-regular energy spacing, performed
in order to investigate the $\Upsilon(6S)$ region. Altogether,
there are 136 energy points~\cite{babar_rb}. In the BaBar paper,
the ISR produced narrow $\Upsilon$ states, i.e., $\Upsilon(1S)$,
$\Upsilon(2S)$, and $\Upsilon(3S)$ states were included in $\Rb$,
but in the data file supplied, their contribution is listed and
can be removed from the data.

The Belle measurement~\cite{belle_rb} was done with the scan data
samples above 10.63~GeV at total 78 data points. The data consist
of one data point of 1.747~fb$^{-1}$ at the peak
$\sqrt{s}=10.869$~GeV; approximately 1~fb$^{-1}$ at each of the 16
energy points between 10.63 and 11.02~GeV; and 50~pb$^{-1}$ at
each of 61 points taken in 5~MeV steps between 10.75 and
11.05~GeV. The non-resonant $q\bar q$ continuum
$(q\in\{u,d,s,c\})$ background is obtained using a 1.03~fb$^{-1}$
data sample taken at $\sqrt{s}=10.52$~GeV. Belle experiment
supplied a data file of $\Rb$ with the ISR produced $\Upsilon(1S)$,
$\Upsilon(2S)$, and $\Upsilon(3S)$ states removed 
(defined as $\Rb^\prime$ in Belle paper~\cite{belle_rb}).

\begin{figure*}[htbp]
 \centering
 \includegraphics[height=6cm]{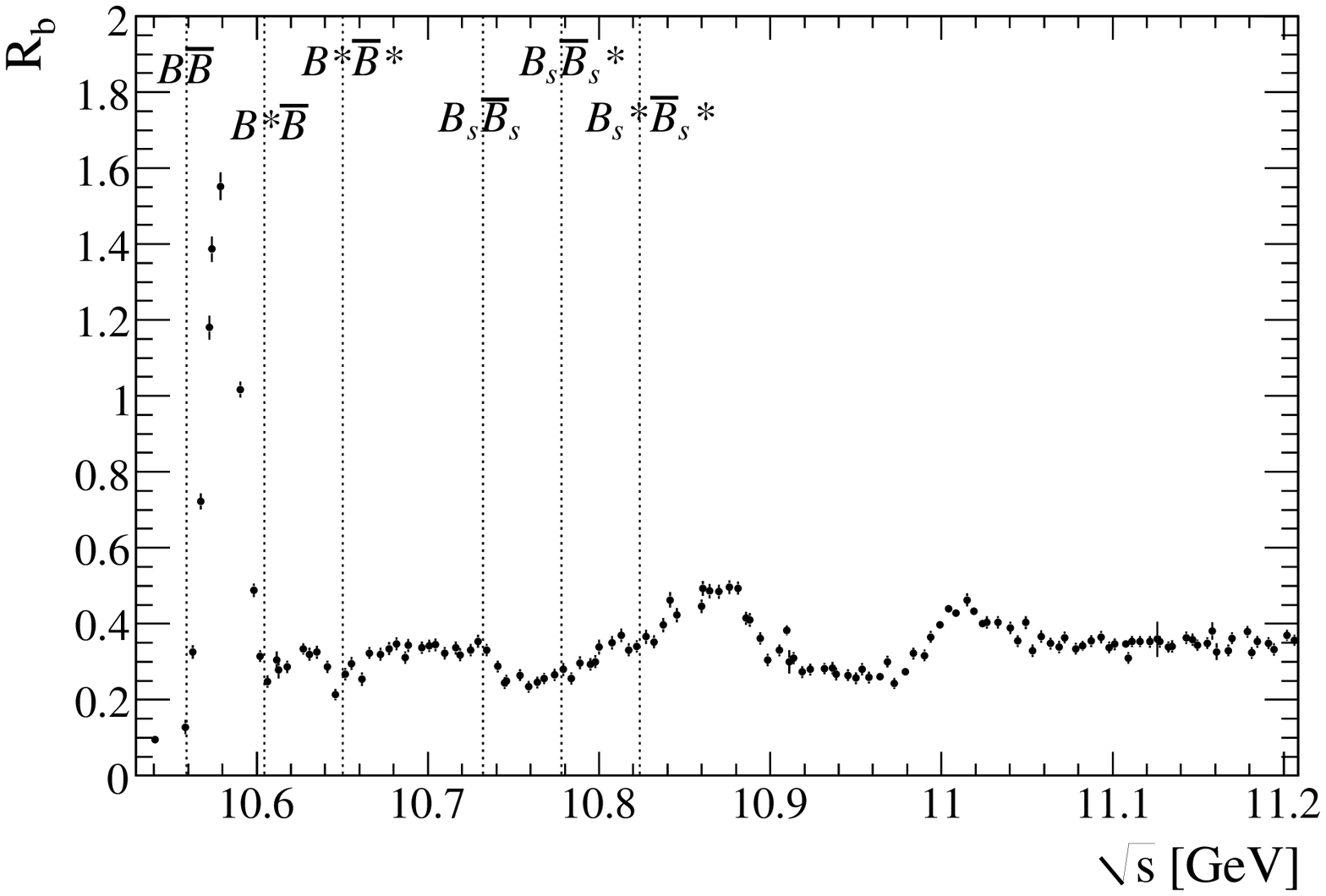}
 \includegraphics[height=6cm]{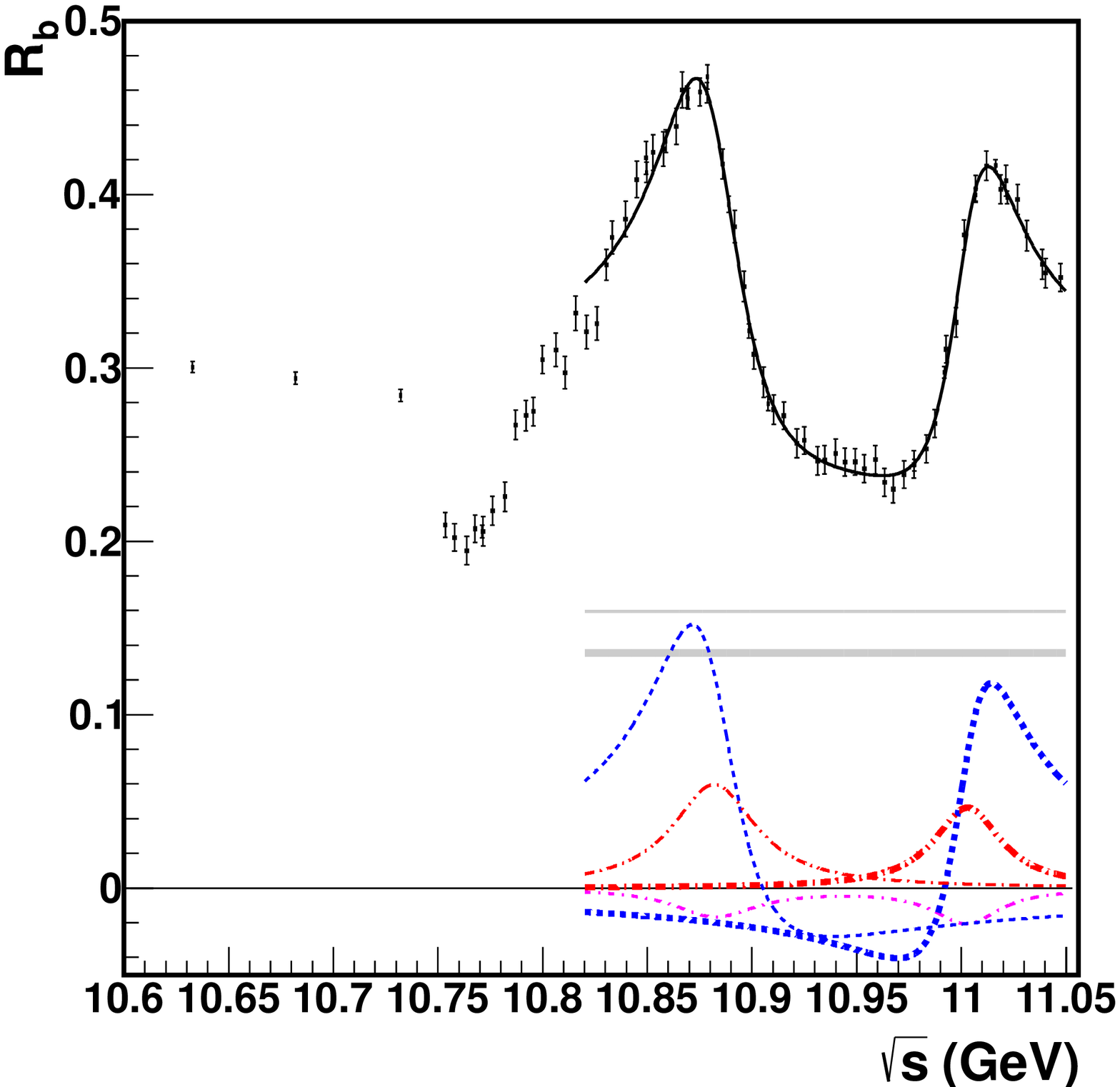}
\caption{$R_b$ data from BaBar~\cite{babar_rb} (left) and
Belle~\cite{belle_rb} (right) experiments. Error bars are
statistical only. The curves are the fit described in
the original paper. } \label{fig:rb}
\end{figure*}

\begin{figure*}[htbp]
 \centering
 \includegraphics[width=6cm,angle=-90]{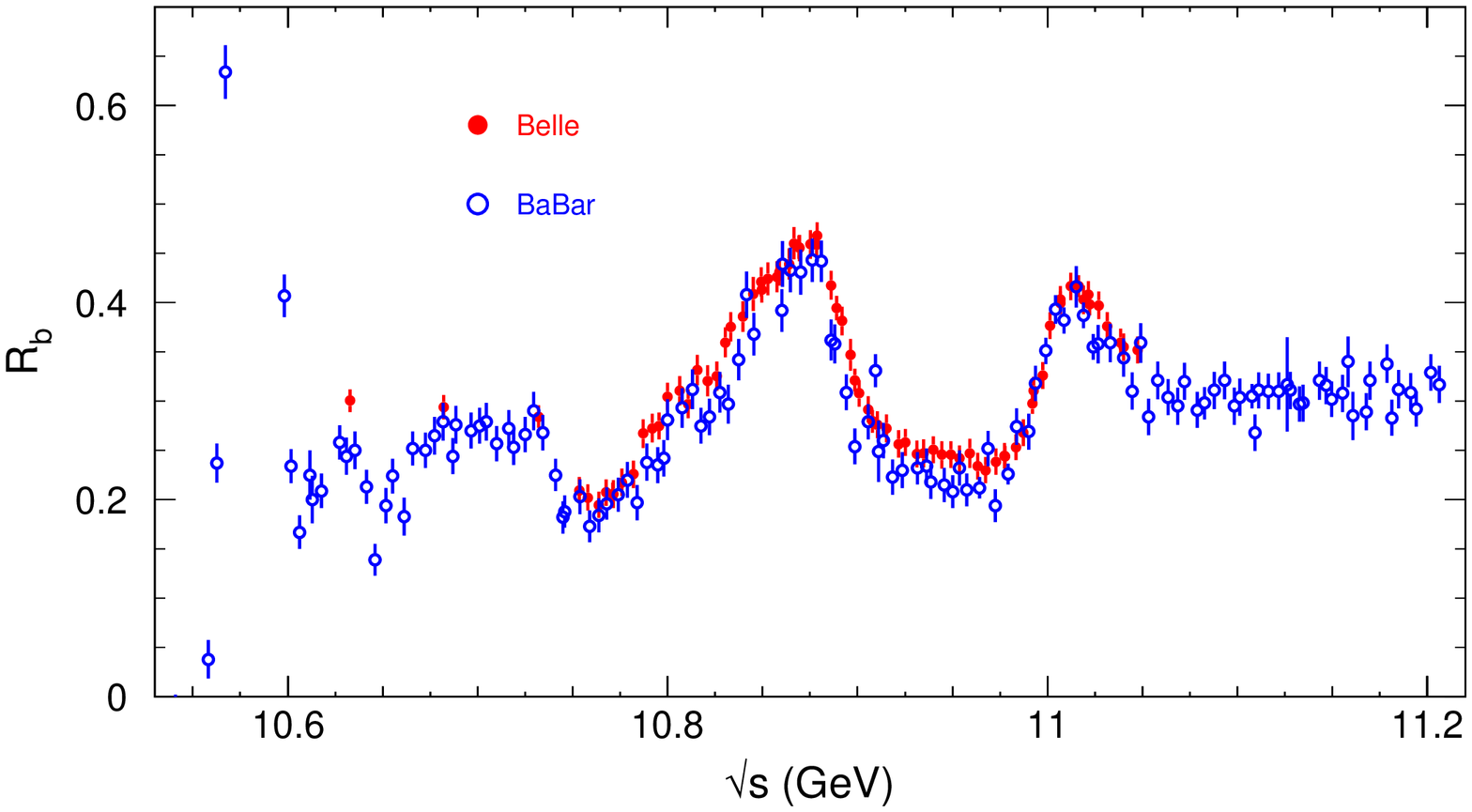} \\
 \includegraphics[width=6.1cm,angle=-90]{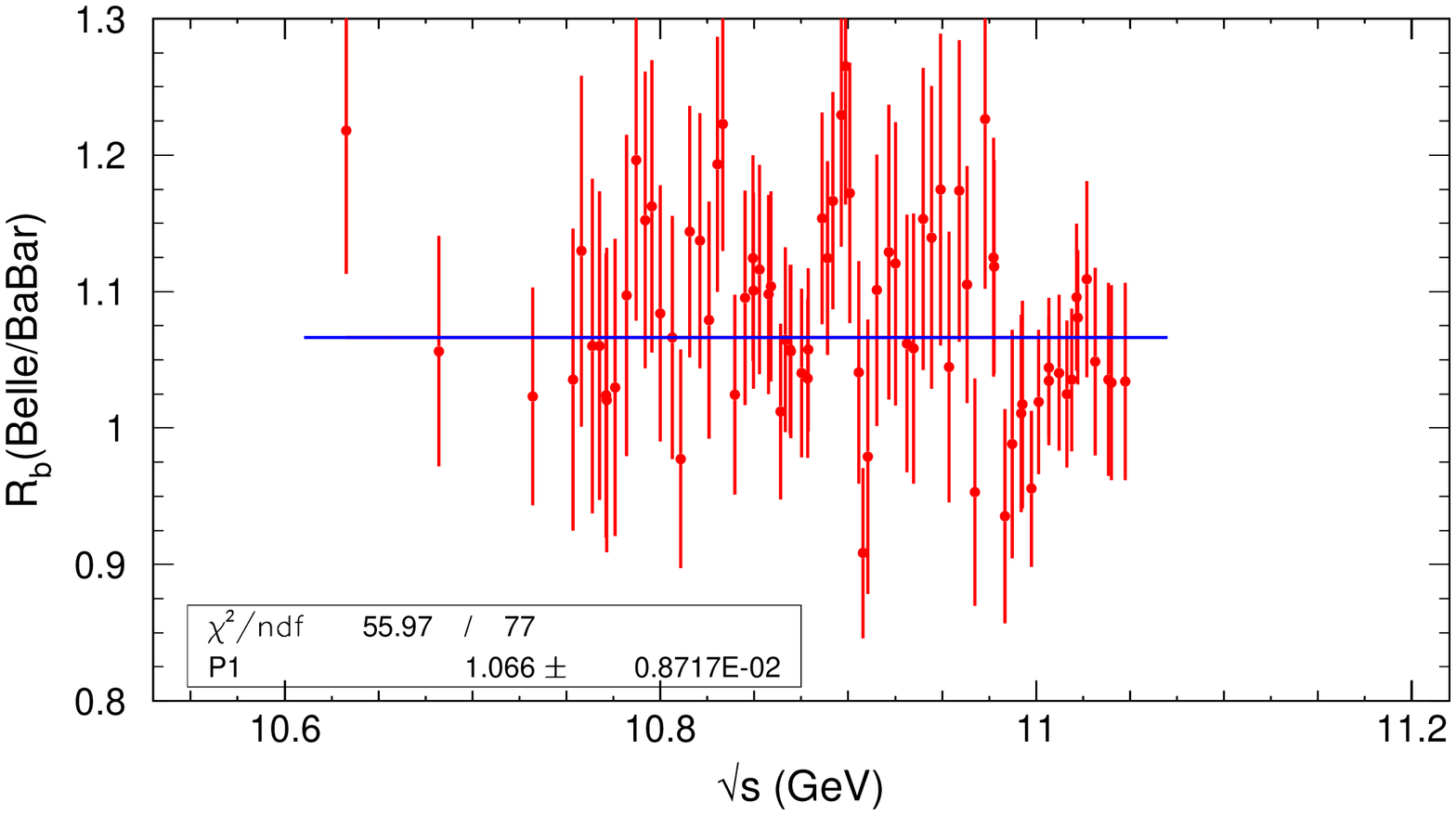}
\caption{Comparison of $R_b$ data from BaBar (open cycles) and
Belle (Red dots) [top panel] experiments and the ratio of $\Rb$
between Belle and BaBar measurements [bottom panel]. Error bars
are combined statistical and systematic errors, and the line is a
fit to the ratio of Belle and BaBar measurements.}
\label{Rb_compare}
\end{figure*}

The BaBar and Belle measurements~\cite{babar_rb,belle_rb} 
are shown in Fig.~\ref{fig:rb}.
Notice that the definitions of $\Rb$ are different in these
papers. After removing the ISR contribution of the narrow
$\Upsilon$ states from BaBar results, the $\Rb$ values and the
comparison between the two experiments are shown in
Fig.~\ref{Rb_compare}. In the following analysis, $\Rb$ refers to
the results after removing the ISR contribution of the narrow
$\Upsilon$ states, $\Rb^{\rm dre}$ refers to the dressed 
cross section after the ISR correction is applied, 
and $\RbB$ refers to the Born cross section
after the ISR and VP corrections are applied.

We can see from Fig.~\ref{Rb_compare} that Belle results are
systematically larger than BaBar measurements. To get the size of
the systematic difference, we calculate the ratio between the
Belle and BaBar measurements in the energy region covered by both
experiments. Figure~\ref{Rb_compare} shows the ratio of the $\Rb$
between Belle and BaBar measurements, the ratios are fitted with a
constant with a good fit quality, $\chi^2/{\rm ndf}=56/77$, 
where ${\rm ndf}$ is the number of degrees of freedom. 
This indicates that the Belle and BaBar measurements differ by a
factor of
\begin{equation}\label{rf}
    f=1.066\pm 0.009,
\end{equation}
which is more than $7\sigma$ from one if they are the same.


\subsection{Combination of Belle and BaBar data}

The BaBar experiment measured the $\Rb$ above 11.1~GeV which is very
flat. This indicates that the bottomonium resonance region has been
passed and the flat continuum region has been reached. At CM energy
well above the open-bottom threshold, $R$ values (the total cross
section of $\EE$ annihilation) and $R_b$ can be calculated with pQCD with 
five different flavors of quarks. And this can be compared with the BaBar
measurement. If we assume the difference in $\Rb$ between Belle
and BaBar can be extrapolated to energy region above 11.1~GeV, by
comparing the expected Belle measurements in this energy region
and the pQCD expectation, we can check the normalization of the
Belle data.

To compare $\Rb$ with pQCD calculation, ISR correction and VP 
correction should be applied to the Belle and BaBar
measurements since pQCD calculates the Born cross sections. Using
the ISR correction factors (point by point correction, average
correction factor $1+\delta\approx 0.901$ above 11.1~GeV) and
VP factors ($\frac{1}{|1-\Pi|^2}\approx 1.076$)
calculated below, a fit to the $\RbB$ from BaBar experiment for CM
energies between 11.10 and 11.21~GeV yields $\RbB=0.316\pm 0.011$
with the error dominated by the common systematic error. 

Assuming Eq.~\eref{rf} applies to $\RbB$ at CM energy above
11.1~GeV for Belle measurement, we extrapolate the Belle
measurement to this energy region so that we would expect 
\begin{equation}\label{rbb_belle}
    \RbB=(0.316\pm 0.011)\times (1.066\pm 0.009)=0.337\pm 0.012.
\end{equation}

Calculating $\RbB$ and the total continuum $R$ values from $udsc$-quarks 
in pQCD according to Eq.~\eref{eq:rqcd}, we find that $\RbB$ is almost 
a constant for CM energy between 11.10~GeV and
11.21~GeV, which is 0.351 with an uncertainty negligible compared
with the experimental measurement; and $R^B_{udsc}$ can
be well parameterized as a linear function of CM energy ($\sqrt{s}$
in GeV) between 10 and 12~GeV, i.e.,
\begin{equation}\label{r}
    R^B_{udsc}=3.5769-4.1249\times 10^{-3}\sqrt{s}.
\end{equation}

\begin{figure*}[htbp]
 \centering
 \includegraphics[height=10.0cm,angle=-90]{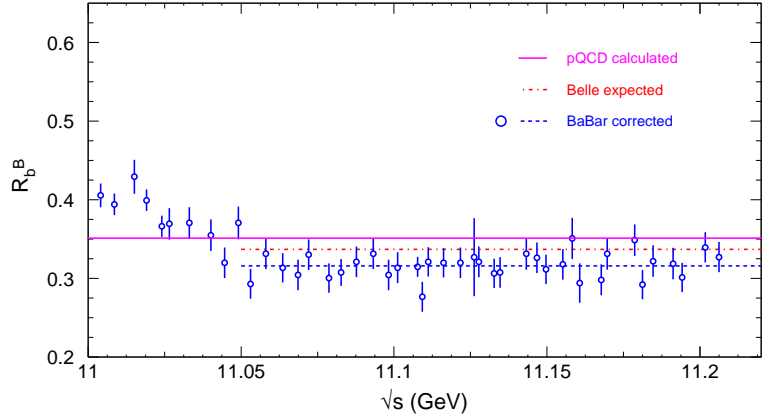}
\caption{The $\RbB$ data from BaBar after ISR and VP 
corrections (open cycles) and the fit with a constant
function (blue dashed line, $\RbB=0.316$) and the expected Belle
results (Red dash-dotted line, $\RbB=0.316\times 1.066=0.337$),
and the pQCD calculation (pink line, $\RbB=0.351$). Error
bars are combined statistical and systematic errors.}
\label{pqcd_compare}
\end{figure*}

\begin{figure*}[htbp]
 \centering
 \includegraphics[height=12cm,angle=-90]{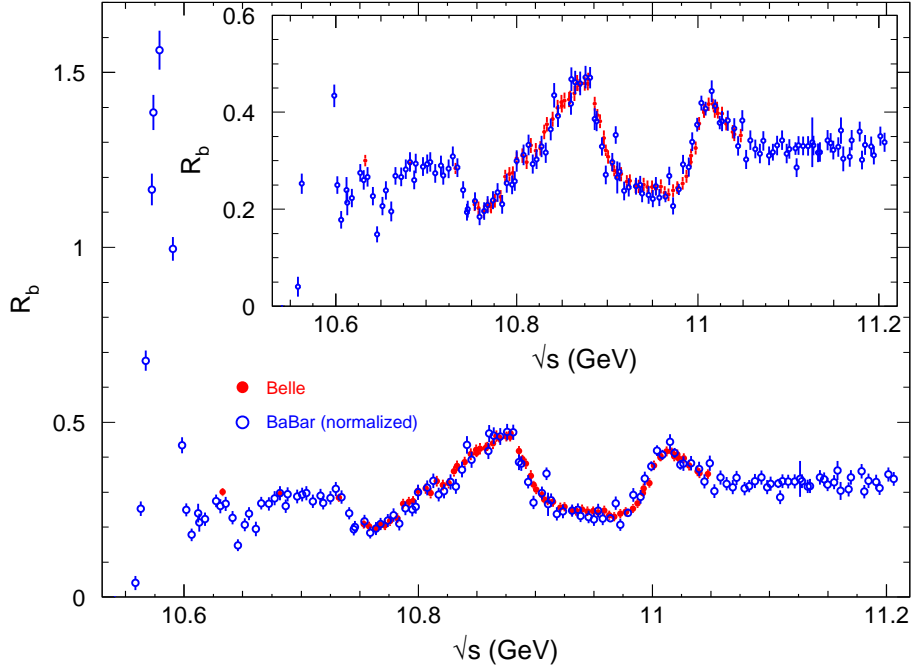}
\caption{Normalized $R_b$ data from BaBar (open cycles) and Belle
(Red dots), which will be treated as a single data set. Error bars
are combined statistical and systematic errors.}
\label{Rb_combine}
\end{figure*}

Figure~\ref{pqcd_compare} shows the comparison between the BaBar
measurements, Belle expected, and the pQCD calculated $\RbB$, we
can find that Belle data agree with pQCD reasonably well (within
about $1\sigma$, the common error of the Belle measurements at high
energy is about $\pm 0.011$, similar to the BaBar measurements)
while the BaBar measurements are about $3\sigma$ lower than pQCD
calculation. As a consequence, we assume BaBar measurement suffers from a
normalization bias, and the Belle measurement is normalized
properly. In the analysis below, we increase the BaBar
measurements by the factor $f$ in Eq.~\eref{rf} and combine them with
the Belle measurements to treat them as a single data set. The
normalized and combined data are shown in Fig.~\ref{Rb_combine}.

In the remainder of this work, the threshold of open-bottom production 
is set to be 10.5585~GeV, larger than the first two energy points 
in BaBar experiment. Therefore, these two data were omitted in 
our analysis.

\subsection{Parametrization of $\Rb$}

To calculate the ISR correction factors, the measured $\Rb$ will
be used as input. To avoid the point-to-point statistical
fluctuation, one may parameterize the line shape with a smooth
curve. There is no known function describing the line shape a
priori, so one may parameterize the line shape with
any possible combination of smooth curves.

\begin{figure*}[htbp]
 \centering
 \includegraphics[height=12cm,angle=-90]{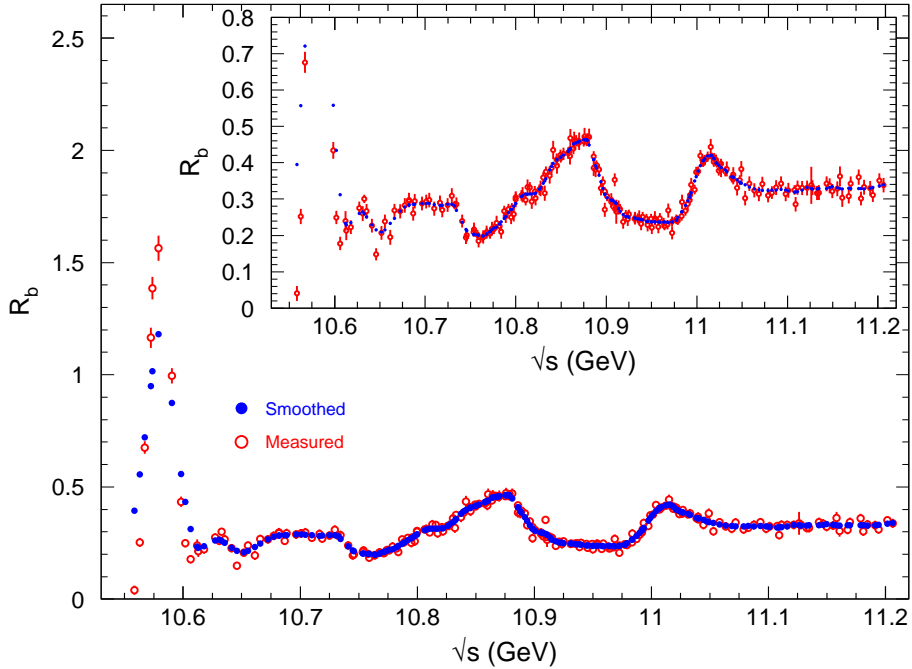}
\caption{Belle and BaBar combined $R_b$ data (red circles with
error bars) and the results after smoothing (blue dots). Error
bars are combined statistical and systematic errors.}
\label{Rb_smooth}
\end{figure*}

We use the ``robust locally weighted regression'' or ``{\sc
lowess}'' method to smooth the experimental measurements. The
principal routine {\sc lowess} computes the smoothed values using
the method described in Ref.~\cite{lowess}. This method works very
well only for slowly varying data, which makes the procedure at the
$\Upsilon(4S)$ region work improperly. As a consequence, we use
the data points directly for $\sqrt{s}<10.66$~GeV and use the
smoothed data for the other data points. Figure~\ref{Rb_smooth}
shows the smoothed $\Rb$, which looks very reasonable.

In the following analysis, we use a straight line to connect two
neighboring points. As these points are after smoothing and the
step is not big, there is no big jump between neighboring points,
so we do not expect significant difference between a straight line
and a smooth curve.

\section{Calculation Procedure}

\subsection{Calculation of ISR correction factors}

We follow the procedure defined in
Eqs.~\eref{isr-factor}, \eref{isr-factor2}, and \eref{isr-factorf} to
calculate the ISR correction factors. 
In doing this for experimental data, we assumed the detection efficiencies 
for $\bbb$ events without ISR and those with different energy of ISR photons 
have been estimated reliably within the quoted systematic uncertainties 
at both BaBar~\cite{babar_rb} and Belle experiment~\cite{belle_rb}.
The iteration is continued until the
difference between two consecutive results is less than 1\% of
the statistical error of the observed $R_b$. 

\begin{figure*}[htbp]
 \centering
 \includegraphics[width=12cm,angle=0]{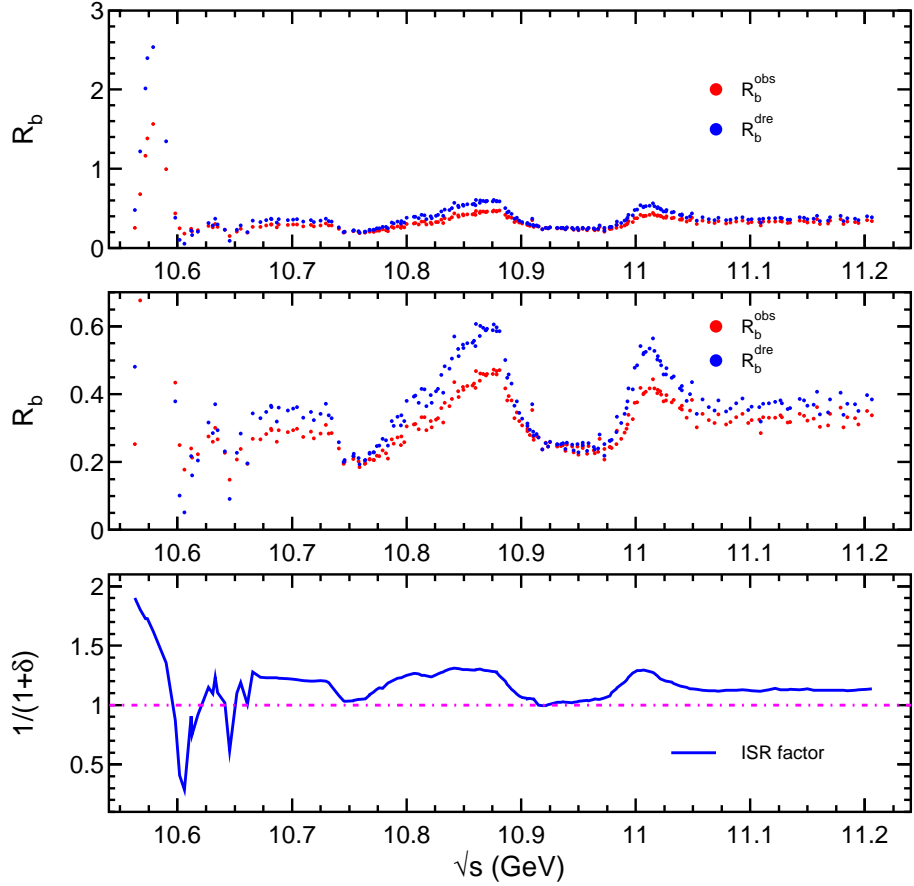}
\caption{Normalized $R_b$ (red in the top two panels) and the ISR
corrected $R_b$ (or $\Rb^{\rm dre}$, blue in the top two
panels), errors are not shown.
The bottom panel shows the ISR correction factor.} \label{isr_plt}
\end{figure*}

In the energy region
where the cross section varies smoothly, the ISR correction
factors become stable after a few iterations while in the
$\Upsilon(4S)$ energy region, due to the rapid change of the cross
section in narrow energy region, the ISR correction factors only
converge to within 1\% after more than ten iterations. We iterate
20 times, and the maximum difference is less than 0.5\% within the
full energy region. Figure~\ref{isr_plt} shows the final ISR 
factors as well as the corrected $\Rb^{\rm dre}$ values.

\subsection{Calculation of vacuum polarization factors}

\begin{figure*}[htbp]
 \centering
 \includegraphics[width=12cm,angle=0]{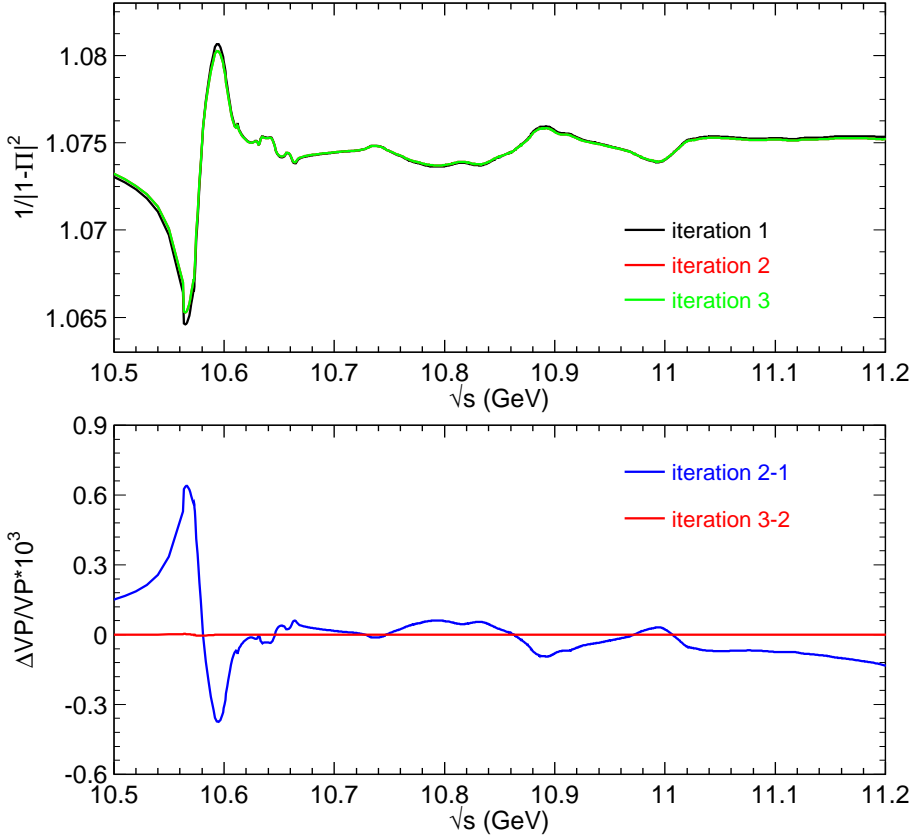}
\caption{The VP factors from 3 iterations (top) and
the difference between two iterations (bottom). Notice that
the difference between the second and the third iterations 
is very small and the curves for the VP factors are almost
indistinguishable.}
\label{vp_bb_compare}
\end{figure*}

Taking $\Rb^{\rm dre}$, the ISR corrected $\Rb$ obtained in previous
subsection, as the approximation of $\RbB$ and adding the
pQCD calculation of the $udsc$-quark contribution to $R$ values
(refer to Eq.~\eref{r}), we calculate the VP
factors in the bottomonium energy region. After obtaining the
VP factor $\frac{1}{|1-\Pi|^2}$, we use $\Rb^{\rm dre}/\frac{1}{
|1-\Pi|^2}$ as input to calculate $\frac{1}{|1-\Pi|^2}$ again and we
iterate this process. It turns out that after three iterations,
the VP factor $\frac{1}{|1-\Pi|^2}$ becomes stable so we
take the values from this round as the final results, and we
obtain $\RbB$ with Eq.~\eref{formula-csf}.
Figure~\ref{vp_bb_compare} shows the VP factors
from this calculation.

\subsection{Estimation of errors}

In the previous two subsections we obtain the ISR correction
factor and VP factor and in turn the Born $\RbB$
via Eq.~(\ref{formula-csf}). During the calculation, however, only
the central values of the observed $R_b$ are used. Instead of 
just scaling the errors of the original measurements by the 
obtained ISR correction factors and the VP 
factors, we perform a toy Monte Carlo sampling to investigate how 
the errors (both statistical and systematic) of the original 
measurements impact the obtained $\RbB$.

At each energy point where the Belle or BaBar
measurement~\cite{babar_rb,belle_rb} was performed, 
we perform 10,000 samplings of the observed
$R_b$ according to a Gaussian distribution for which the mean value and
standard deviation are the central value and statistical error of
the observed $R_b$, respectively. These samples will be used to 
estimate the statistical errors of the deduced quantities.  
In addition, the uncommon
systematic errors are also added to the samples in the same way. 
For the common systematic error, the same error is added to each 
sample at all energy points in Belle or BaBar experiment. 
These samples with both statistical and systematic errors considered 
will yield the total errors of the deduced quantities. In the 
energy regions $0.36<\sqrt{s}<2.0$~GeV and $3.7<\sqrt{s}<5.0$~GeV, 
the data from the PDG compilation~\cite{PDG,experiments} and BES 
collaboration~\cite{Bai:1999pk,Bai:2001ct} are assumed to be completely 
correlated when we perform the sampling.

With each sample as input, we repeat the calculation described in the
previous two subsections to obtain the ISR correction factor,
VP factor, and $\RbB$ for this sample. Finally a distribution 
of $\RbB$ at each energy point is observed, so are the ISR
correction factor and the VP factor. We find that
these distributions also satisfy Gaussian distribution well so the
fitted mean and standard deviation are taken as the central value
and error of the corresponding quantities,
respectively. The covariances of the distributions of Born and 
dressed cross sections at different energy points are also available 
in the supplemental material~\cite{supple}. These covariances are useful in 
calculations where $R_b^{\rm B}$ or $R_b^{\rm dre}$ are inputs such 
as extracting the resonant parameters of the $Y(10750)$, $\Upsilon(5S)$, 
and $\Upsilon(6S)$ by fitting $R_b^{\rm dre}$.

Recall that we smoothed the observed $R$ values using the ``Lowess" 
method before we calculated the dressed and Born ones. In principle, one can 
use different methods to smooth the data, which will result in uncertainty 
of the final results. We test another smoothing method, ``Smoothing spline"~\cite{smo_spline}, 
and find that such uncertainty is negligible when compared with the 
original errors.

\section{Final results on $\RbB$}

After all the above operations, we obtain the $\RbB$ as well as
its total uncertainty from the combined BaBar and Belle measurements as
shown in Fig.~\ref{rbb_final} and the Table in the Appendix. 
We find that the $\RbB$ values are very different from the $\Rb$
reported from the original publications~\cite{babar_rb,belle_rb}
and the differences are energy dependent. Common features are that
the peaks are even higher and the valleys become deeper, the two
dips at the $B\bar{B^*}+c.c.$ and $B^*\bar{B^*}$ thresholds are
more significant, the peaks corresponding to the $\Upsilon(5S)$
and $\Upsilon(6S)$ increase significantly, and there is a 
prominent dip at 10.75~GeV.

\begin{figure*}[tbp]
 \centering
 \includegraphics[height=12cm,angle=-90]{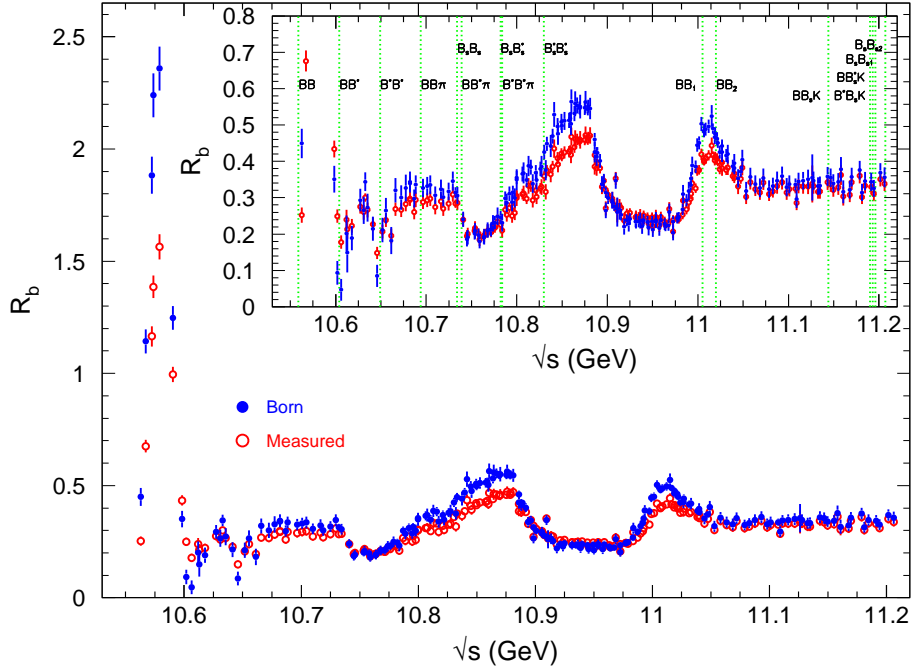}
\caption{Comparison of measured $R_b$ (open cycles) and Born
$\RbB$ (solid dots). The error bars are the combined statistical
and systematic errors. The dotted vertical lines are the thresholds 
for bottom meson productions.} \label{rbb_final}
\end{figure*}

The total $R$ value corresponding to the production of $udscb$ quarks
can be obtained directly by adding the $\RbB$ to the $udsc$-quark 
contribution calculated from pQCD, as indicated in Eq.~\eref{r}.

\section{Summary and discussions}

From the BaBar and Belle measurements of the observed $\EE\to
\bbb$ cross sections, we do ISR correction to obtain the dressed
$\EE\to \bbb$ cross sections from 10.56 to 11.21~GeV. These
dressed cross sections are the right ones to be used to
determine the resonant parameters of the vector bottomonium states. 
Together with the $R$ values measured at other
energy points and the $R$ values calculated with pQCD, we
calculate the VP factors. By applying VP correction, we obtain the Born cross section
of $\EE\to \bbb$ from threshold to 11.21~GeV. These cross sections can
be used for all the calculations related to the photon propagator,
such as $a_\mu$, the $\mu$ anomalous magnetic moment, and $\alpha(s)$,
the running coupling constant of QED~\cite{Davier,Teubner}.

In the following parts of this section, we discuss the usage of
the data obtained in this study.

\subsection{Vacuum polarization}

The VP factors have been calculated by many
groups~\cite{radcorr,Berends_vp,ignatov,vacuum,HMNT}, with the
experimental data and various theoretical inputs when the data are
not available or less precise. Different techniques on how to
handle the discrete data points and how to correct possible bias
in data were developed. All these different treatments yield
very similar results on hadronic contribution to $a_\mu$ and on
the running of the $\alpha$ at $M_{Z^0}^2$, which indicates that
the methods are all essentially applicable with current precision
of data.

\begin{figure*}[htbp]
\includegraphics[width=15.cm]{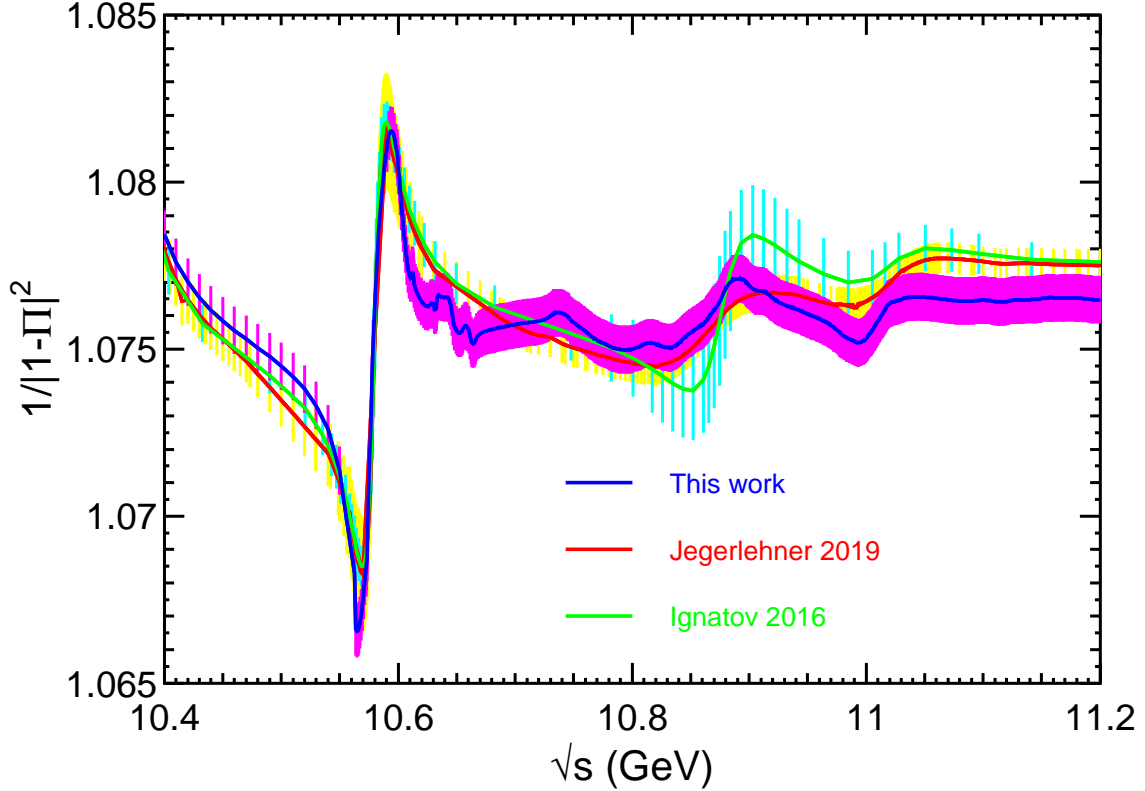}
\caption{\label{fig:vp_compare} VP factors in
the bottomonium energy region and the comparison with previous
calculations~\cite{vacuum,ignatov}. The solid lines are the central 
values and the error bars or bands show the uncertainties.}
\end{figure*}

Previous calculations of the VP factors in the bottomonium energy 
region used either the resonant parameters of the
$\Upsilon(4S)$, $\Upsilon(5S)$, and $\Upsilon(6S)$ reported by
previous experiments~\cite{CUSB,CLEO} which are very crude~\cite{ignatov,vacuum}
or the experimental data from previous experiments~\cite{CUSB,CLEO} which gave 
the observed cross sections~\cite{HMNT}. 
We recalculate the VP factors by using $\RbB$ obtained 
in this analysis based on high precision data from BaBar and Belle
experiments~\cite{babar_rb,belle_rb}, with the ISR correction and
VP factors properly considered. Although these new
data have little effect on the VP factors far from
the bottomonium energy region, they do change the 
VP factors in the bottomonium energy region as is shown
in Fig.~\ref{fig:vp_compare}. The difference between this and the
previous calculations~\cite{vacuum,ignatov} is visible 
at some energies although all the calculations agree within errors.

\subsection{Bottomonium spectroscopy}

There are very clear structures in $\RbB$ distribution shown in 
Fig.~\ref{rbb_final}. From low to high energy, we identify
the $\Upsilon(4S)$ at 10.58~GeV, dips due to
$B\bar{B^*}+c.c.$ and $B^*\bar{B^*}$ thresholds at 
10.61 and 10.65~GeV, respectively, a dip at 10.75~GeV
that may correspond to the $Y(10750)$~\cite{Y10750},
and the $\Upsilon(5S)$ and $\Upsilon(6S)$ at 10.89
and 11.02~GeV, respectively.

The observed $\Rb$ values were used to extract the resonant
parameters of the $\Upsilon(5S)$ and $\Upsilon(6S)$ in the
BaBar~\cite{babar_rb} and Belle~\cite{belle_rb} publications. As
the ISR correction effect is significant and is energy dependent,
this suggests that the fit results are not reliable.
To avoid the dip at around 10.75~GeV, both BaBar and Belle fitted 
data above 10.80~GeV only. A recent study of $\EE\to \pp\Upsilon$
revealed a new state, the $Y(10750)$, with a mass of 
$(10752.7\pm 5.9^{+0.7}_{-1.1})$~MeV/$c^2$ and width 
$(35.5^{+17.6}_{-11.3}{}^{+3.9}_{-3.3})$~MeV~\cite{Y10750}, at 
exactly the position of the dip in $\RbB$. This indicates that
the dip is very likely to be produced by the interference between a 
Breit-Wigner function and a smooth background component.

We do a least-square fit to the dressed $\EE\to \bbb$ cross sections 
($\sigma^{\rm dre}=\frac{\sigma^B}{|1-\Pi|^2}$) above 10.68~GeV
with the coherent sum of a continuum amplitude (proportional to $1/\sqrt{s}$)
and three Breit-Wigner functions with constant widths representing the
structures at 10.75, 10.89, and 11.02~GeV. The Breit-Wigner function is
$$
{\rm BW}=e^{i\phi}\frac{\sqrt{12\pi\Gamma_{\EE}\Gamma}}{s-m^2+im\Gamma},
$$
where $m$, $\Gamma$, $\Gamma_{\EE}$, and $\phi$ are the mass, total width,
electronic partial width of the resonance, and the relative
phase between the resonance and the real continuum amplitude, respectively, 
and they are all free parameters in the fits.

Eight sets of solutions are found from the fit~\cite{zhuk}, 
with identical total fit curve, identical fit quality
($\chi^2=274$ with 188 data points and 13 free parameters), and identical 
masses and widths for the same resonance, but with significantly different 
$\Gamma_{\EE}$ and $\phi$.

\begin{figure*}[htbp]
 \centering
 \includegraphics[height=12cm,angle=-90]{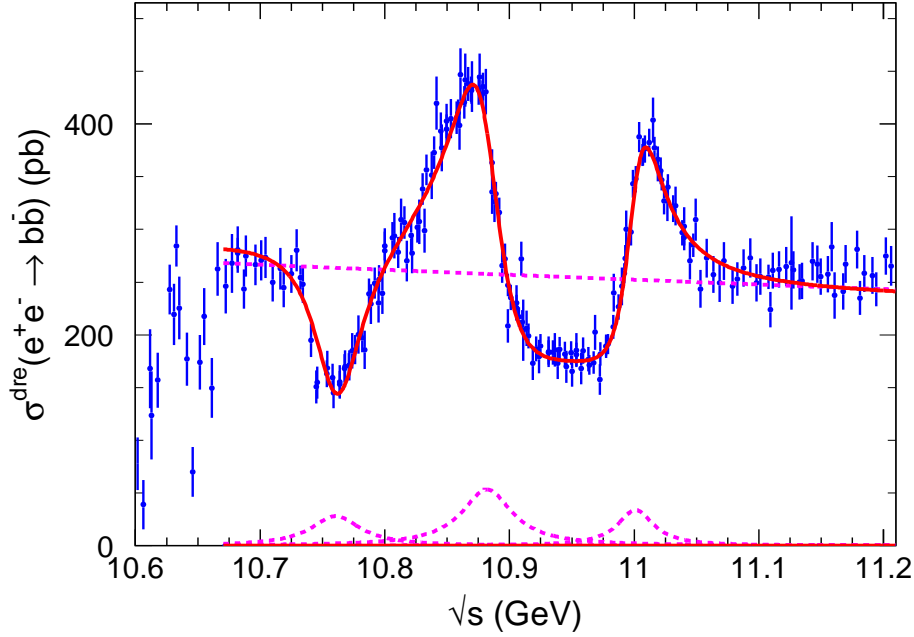}
\caption{Fit to the dressed cross sections with coherent sum of a continuum
amplitude and three Breit-Wigner functions. The solid curve is the total fit, and
the dashed ones for each of the four components from Sol.~1 in 
Table~\ref{tab:res_param}. The magnitudes of these components are different
in different solutions.} \label{rb_fit_sol1}
\end{figure*}

Figure~\ref{rb_fit_sol1} shows one of the solutions of the fit, and
Table~\ref{tab:res_param} lists the resonant parameters from 
8 solutions of the fit. The masses and widths of the resonances 
agree with those from Ref.~\cite{Y10750} but with improved precision
because of the much better measurements used in this work compared
with those in exclusive $\EE\to \pp\Upsilon$ analyses. The $\Gamma_{\EE}$
values determined from this study allow us to extract the branching fractions of
$\pp\Upsilon$ of these resonances by combining the information reported
in Ref.~\cite{Y10750}, and to understand the nature of these vector 
states~\cite{zhangal,maiani,zhongxh,wangzg,liangwh}.

\begin{table}[htbp]
\caption{Resonant parameters from the fit to dressed cross sections. There are 
8 solutions with identical fit quality, and the masses and widths of the resonances
are identical in all the solutions. The uncertainties are combined statistical and systematic uncertainties in experimental measurements.}
    \label{tab:res_param}
    \centering
    \begin{tabular}{C{2cm}|C{2.5cm}C{2.5cm}C{2.5cm}C{2.5cm}}
    \hline\hline
 Solution & Parameter      &   $Y(10750)$    &  $\Upsilon(5S)$   & $\Upsilon(6S)$   \\\hline
\multirow{2}*{1--8} 
    & Mass (MeV/$c^2$)     &  $10761\pm 2$   &   $10882\pm 1$    &  $11001\pm 1$    \\
    & Width (MeV)          &  $48.5\pm 3.0$  &   $49.5\pm 1.5$   &  $35.1\pm 1.2$   \\\hline
\multirow{2}*{1} 
  & $\Gamma_{\EE}$ (eV)    &  $10.7\pm 0.9$  &   $21.3\pm 1.0$   &  $9.8\pm 0.5$    \\
  &    $\phi$ (degree)     &   $260\pm 3$    &    $144\pm 2$     &   $34\pm 3$      \\\hline
\multirow{2}*{2} 
  & $\Gamma_{\EE}$ (eV)    &  $11.1\pm 0.9$  &   $24.8\pm 1.3$   &   $307\pm 9$    \\
  &    $\phi$ (degree)     &   $270\pm 3$    &    $164\pm 2$     &   $280\pm 1$     \\\hline
\multirow{2}*{3} 
  & $\Gamma_{\EE}$ (eV)    &  $12.6\pm 1.1$  &    $479\pm 14$    &  $11.5\pm 0.6$   \\
  &      $\phi$ (degree)   &   $295\pm 3$    &    $254\pm 1$     &   $3\pm 3$       \\\hline
\multirow{2}*{4} 
  & $\Gamma_{\EE}$ (eV)    &  $13.0\pm 1.1$  &    $558\pm 19$    &   $363\pm 13$    \\
  &       $\phi$ (degree)  &   $296\pm 3$    &    $274\pm 1$     &   $249\pm 1$     \\\hline
\multirow{2}*{5} 
  & $\Gamma_{\EE}$ (eV)    &   $324\pm 24$   &   $23.7\pm 1.2$   &  $10.0\pm 0.5$    \\
  &      $\phi$ (degree)   &   $265\pm 1$    &    $129\pm 2$     &   $26\pm 3$      \\\hline
\multirow{2}*{6} 
  & $\Gamma_{\EE}$ (eV)    &   $336\pm 27$   &   $27.6\pm 1.6$   &   $314\pm 10$    \\
  &      $\phi$ (degree)   &   $275\pm 1$    &    $149\pm 2$     &   $272\pm 1$     \\\hline
\multirow{2}*{7} 
  & $\Gamma_{\EE}$ (eV)    &   $380\pm 32$   &    $534\pm 18$    &  $11.8\pm 0.6$   \\
  &      $\phi$ (degree)   &   $291\pm 1$    &    $239\pm 1$     &   $355\pm 3$     \\\hline
\multirow{2}*{8} 
  & $\Gamma_{\EE}$ (eV)    &   $394\pm 34$   &    $622\pm 25$    &   $370\pm 14$    \\
  &      $\phi$ (degree)   &   $301\pm 1$    &    $259\pm 2$     &   $241\pm 1$     \\
    \hline\hline
    \end{tabular}
\end{table}

In this analysis, we assumed that all the resonances are Breit-Wigner functions 
with constant widths and the continuum term is a smooth curve in the
full energy region and they interfere with each other completely. In fact,
the $\RbB$ or the total cross section has contributions from different modes,
including open bottom and hidden bottom final states, the
parametrization of the line shape should be very complicated due
to the coupled-channel effect~\cite{Tornqvist:1984fx} and the
presence of many open bottom thresholds, $B\bar{B}$,
$B\bar{B^*}+c.c.$, $B^*\bar{B^*}$, $B_s\bar{B_s}$,
$B_s\bar{B_s^*}+c.c.$, $B_s^*\bar{B_s^*}$, $B\bar{B_1}+c.c.$, 
$B^*\bar{B_0}+c.c.$, $B^*\bar{B_1}+c.c.$, $B\bar{B_2}+c.c.$, and
so on, and even $\pi Z_b(10610)$, $\pi Z_b(10650)$. The situation
becomes somewhat simpler in a single final state like
$\pp\Upsilon(nS)$ ($n=1,\, 2,\, 3$)~\cite{belle_rb} and $\pp
h_b(mP)$ ($m=1,\, 2$)~\cite{pipihb} although the intermediate structure in the
three-body final state is also complicated. The results from these
fits may change dramatically by including more information on each exclusive
mode. 

We also attempt to add one more Breit-Wigner function to fit the cross sections,
the fit quality improves slightly with a state at $m=(10848\pm 9)$~MeV/$c^2$ 
with a width of $(28\pm 14)$~MeV, a state at $m=(10831\pm 1)$~MeV/$c^2$ 
with a width of $(14\pm 5)$~MeV, or a state at $m=(11065\pm 23)$~MeV/$c^2$ 
with a width of $(73\pm 37)$~MeV. In all these cases, the significance of 
the additional state is less than $4\sigma$. 

The data obtained in this analysis can be used to extract 
resonant parameters of these states if a better parametrization
of the cross sections is developed.

\subsection{Search for the production of invisible particles}

The experimentally observed $\EE\to \MM$ cross section, 
with radiative correction is expressed as
\begin{equation}\label{mupair}
\sigma(s)(\EE\to \MM) = \int^{x_m}_0 \frac{4\pi\alpha^2}{3s(1-x)}  
\frac{F(x,s)}{|1-\Pi(s(1-x))|^2} dx,
\end{equation}
where $F(x,s)$ is expressed in Eq.~(\ref{FKuraev}), 
$x_m=1-s_m/s$, with $\sqrt{s_m}$ the required minimum invariant mass 
of the $\mu$ pair in the event selection. 
Here the mass of $\mu$ is neglected for charm and 
beauty factories. In Eq.~(\ref{mupair}), the cross section 
is calculated by QED without any ambiguity except the 
vacuum polarization $\Pi(s)$ which is expressed by 
Eqs.~(\ref{defdeltav}--\ref{dspnrelationb}), with the hadronic 
contribution depending on the experimental measured data as input. 

If the $\EE\to \MM$ cross section is measured to a high precision, 
then $\Pi(s)$ can be obtained. Thus $\Pi(s)$ is measured 
from the experiment directly~\cite{ref-ask-1-kloe}. It can then be compared with 
that calculated by Eqs.~(\ref{defdeltav}--\ref{dspnrelationb}).
This provides a test of QED at high luminosity flavor factories.
In Eq.~(\ref{defdeltav}), the leptonic term $\Pi_l(s,m^2)$ 
is expressed in terms of the QED fine structure constant $\alpha(0)$ 
and lepton masses, which are all known to very high accuracy,
while the hadronic term $\Pi_h(s)$ must be evaluated with 
Eq.~(\ref{dspnrelationb}) with the input of experimental 
measured hadronic cross sections. It is seen in Eq.~(\ref{dspnrelationb})  that
$\Pi_h(s)$ is most sensitive to the hadronic cross section
at energies close to $s$, which can be measured 
with the same experiment. Therefore such a test of QED can 
be performed with two sets of data, $\EE\to \mu^+\mu^-$ and 
$\EE\to {\rm hadrons}$, collected within the same experiment.  
Any discrepancy would mean that there are missing
hadronic final states, or even a final state that
escaped detection, or is invisible by current detection technology. 
This provides a test of QED and search for new physics.
We propose for it to be included in physics goals in 
future high-luminosity frontier physics.

\acknowledgments

This work is supported in part by National Natural Science Foundation 
of China (NSFC) under contract Nos. 11521505, 11475187, and 11375206; 
Key Research Program of Frontier Sciences, CAS, Grant No. QYZDJ-SSW-SLH011; 
the CAS Center for Excellence in Particle Physics (CCEPP); and the 
Munich Institute for Astro- and Particle Physics (MIAPP) which is funded 
by the Deutsche Forschungsgemeinschaft (DFG, German Research Foundation) 
under Germany's Excellence Strategy-EXC-2094-390783311.


\clearpage

\appendix

\section{Final results of Born cross section $R_b^{\rm B}$ and dressed cross section $R_b^{\rm{dre}}$, together with
the ISR correction factors and vacuum polarization factors. The first error of $R_b^{\rm B}$ and $R_b^{\rm{dre}}$ is statistical and the second one is systematic. The errors of ISR factor and VP factor are combined statistical and systematic errors.}

\begin{center}

\begin{longtable}{C{1.6cm}C{4.25cm}C{4.25cm}C{2.75cm}C{2.75cm}}
  \hline\hline
  $\sqrt{s}$ (GeV) & $R_b^{\rm B}$& $R_b^{\rm{dre}}$
          &$1/(1+\delta)$
          & $1/|1-\Pi|^2$\\\hline
  \endhead
  
  \hline\hline
  \endfoot
10.5628 & 0.4498 $\pm$ 0.0341 $\pm$ 0.0155 & 0.4799 $\pm$ 0.0364 $\pm$ 0.0164 & 1.9009 $\pm$ 0.0000 & 1.0670 $\pm$ 0.0008 \\ 
10.5673 & 1.1436 $\pm$ 0.0399 $\pm$ 0.0308 & 1.2199 $\pm$ 0.0425 $\pm$ 0.0325 & 1.8044 $\pm$ 0.0074 & 1.0668 $\pm$ 0.0008 \\ 
10.5723 & 1.8835 $\pm$ 0.0598 $\pm$ 0.0487 & 2.0122 $\pm$ 0.0642 $\pm$ 0.0516 & 1.7273 $\pm$ 0.0063 & 1.0683 $\pm$ 0.0008 \\ 
10.5738 & 2.2419 $\pm$ 0.0714 $\pm$ 0.0574 & 2.3981 $\pm$ 0.0760 $\pm$ 0.0610 & 1.7294 $\pm$ 0.0126 & 1.0697 $\pm$ 0.0008 \\ 
10.5788 & 2.3593 $\pm$ 0.0700 $\pm$ 0.0598 & 2.5363 $\pm$ 0.0750 $\pm$ 0.0642 & 1.6218 $\pm$ 0.0082 & 1.0750 $\pm$ 0.0008 \\ 
10.5903 & 1.2485 $\pm$ 0.0372 $\pm$ 0.0323 & 1.3496 $\pm$ 0.0402 $\pm$ 0.0350 & 1.3543 $\pm$ 0.0121 & 1.0809 $\pm$ 0.0008 \\ 
10.5983 & 0.3508 $\pm$ 0.0326 $\pm$ 0.0119 & 0.3791 $\pm$ 0.0353 $\pm$ 0.0128 & 0.8728 $\pm$ 0.0438 & 1.0809 $\pm$ 0.0008 \\ 
10.6018 & 0.0943 $\pm$ 0.0289 $\pm$ 0.0059 & 0.1019 $\pm$ 0.0312 $\pm$ 0.0063 & 0.4002 $\pm$ 0.1016 & 1.0797 $\pm$ 0.0008 \\ 
10.6063 & 0.0465 $\pm$ 0.0270 $\pm$ 0.0053 & 0.0501 $\pm$ 0.0291 $\pm$ 0.0057 & 0.2732 $\pm$ 0.1383 & 1.0779 $\pm$ 0.0008 \\ 
10.6118 & 0.2022 $\pm$ 0.0444 $\pm$ 0.0079 & 0.2178 $\pm$ 0.0478 $\pm$ 0.0085 & 0.8942 $\pm$ 0.1092 & 1.0771 $\pm$ 0.0008 \\ 
10.6128 & 0.1488 $\pm$ 0.0498 $\pm$ 0.0083 & 0.1603 $\pm$ 0.0536 $\pm$ 0.0090 & 0.7332 $\pm$ 0.1801 & 1.0771 $\pm$ 0.0008 \\ 
10.6178 & 0.1892 $\pm$ 0.0301 $\pm$ 0.0082 & 0.2037 $\pm$ 0.0324 $\pm$ 0.0089 & 0.9110 $\pm$ 0.0810 & 1.0765 $\pm$ 0.0008 \\ 
10.6273 & 0.2938 $\pm$ 0.0263 $\pm$ 0.0102 & 0.3162 $\pm$ 0.0283 $\pm$ 0.0110 & 1.1466 $\pm$ 0.0401 & 1.0763 $\pm$ 0.0008 \\ 
10.6308 & 0.2648 $\pm$ 0.0331 $\pm$ 0.0102 & 0.2850 $\pm$ 0.0356 $\pm$ 0.0110 & 1.0919 $\pm$ 0.0638 & 1.0762 $\pm$ 0.0008 \\ 
10.6328 & 0.3440 $\pm$ 0.0067 $\pm$ 0.0228 & 0.3702 $\pm$ 0.0072 $\pm$ 0.0245 & 1.2330 $\pm$ 0.0444 & 1.0763 $\pm$ 0.0008 \\ 
10.6353 & 0.2728 $\pm$ 0.0335 $\pm$ 0.0123 & 0.2937 $\pm$ 0.0360 $\pm$ 0.0133 & 1.0989 $\pm$ 0.0658 & 1.0766 $\pm$ 0.0008 \\ 
10.6413 & 0.2145 $\pm$ 0.0293 $\pm$ 0.0082 & 0.2310 $\pm$ 0.0315 $\pm$ 0.0088 & 1.0136 $\pm$ 0.0666 & 1.0765 $\pm$ 0.0008 \\ 
10.6458 & 0.0848 $\pm$ 0.0287 $\pm$ 0.0048 & 0.0912 $\pm$ 0.0308 $\pm$ 0.0051 & 0.6015 $\pm$ 0.1485 & 1.0761 $\pm$ 0.0008 \\ 
10.6518 & 0.2120 $\pm$ 0.0297 $\pm$ 0.0103 & 0.2279 $\pm$ 0.0319 $\pm$ 0.0111 & 1.0931 $\pm$ 0.0681 & 1.0754 $\pm$ 0.0008 \\ 
10.6553 & 0.2637 $\pm$ 0.0312 $\pm$ 0.0103 & 0.2836 $\pm$ 0.0335 $\pm$ 0.0111 & 1.1868 $\pm$ 0.0602 & 1.0755 $\pm$ 0.0008 \\ 
10.6613 & 0.1820 $\pm$ 0.0335 $\pm$ 0.0086 & 0.1957 $\pm$ 0.0360 $\pm$ 0.0093 & 0.9961 $\pm$ 0.0892 & 1.0755 $\pm$ 0.0008 \\ 
10.6659 & 0.3194 $\pm$ 0.0255 $\pm$ 0.0116 & 0.3434 $\pm$ 0.0275 $\pm$ 0.0125 & 1.2768 $\pm$ 0.0358 & 1.0754 $\pm$ 0.0008 \\ 
10.6724 & 0.3059 $\pm$ 0.0207 $\pm$ 0.0105 & 0.3290 $\pm$ 0.0223 $\pm$ 0.0113 & 1.2326 $\pm$ 0.0118 & 1.0754 $\pm$ 0.0008 \\ 
10.6774 & 0.3236 $\pm$ 0.0218 $\pm$ 0.0102 & 0.3480 $\pm$ 0.0235 $\pm$ 0.0110 & 1.2307 $\pm$ 0.0104 & 1.0755 $\pm$ 0.0008 \\ 
10.6819 & 0.3405 $\pm$ 0.0225 $\pm$ 0.0101 & 0.3662 $\pm$ 0.0241 $\pm$ 0.0109 & 1.2306 $\pm$ 0.0128 & 1.0755 $\pm$ 0.0008 \\ 
10.6820 & 0.3364 $\pm$ 0.0040 $\pm$ 0.0148 & 0.3618 $\pm$ 0.0043 $\pm$ 0.0159 & 1.2306 $\pm$ 0.0128 & 1.0755 $\pm$ 0.0008 \\ 
10.6869 & 0.2975 $\pm$ 0.0209 $\pm$ 0.0103 & 0.3200 $\pm$ 0.0225 $\pm$ 0.0110 & 1.2273 $\pm$ 0.0130 & 1.0756 $\pm$ 0.0008 \\ 
10.6884 & 0.3357 $\pm$ 0.0230 $\pm$ 0.0104 & 0.3611 $\pm$ 0.0247 $\pm$ 0.0112 & 1.2244 $\pm$ 0.0138 & 1.0756 $\pm$ 0.0008 \\ 
10.6964 & 0.3273 $\pm$ 0.0206 $\pm$ 0.0108 & 0.3521 $\pm$ 0.0221 $\pm$ 0.0116 & 1.2215 $\pm$ 0.0122 & 1.0757 $\pm$ 0.0008 \\ 
10.7009 & 0.3321 $\pm$ 0.0213 $\pm$ 0.0108 & 0.3573 $\pm$ 0.0229 $\pm$ 0.0116 & 1.2181 $\pm$ 0.0129 & 1.0757 $\pm$ 0.0008 \\ 
10.7044 & 0.3357 $\pm$ 0.0221 $\pm$ 0.0107 & 0.3611 $\pm$ 0.0237 $\pm$ 0.0115 & 1.2138 $\pm$ 0.0114 & 1.0757 $\pm$ 0.0008 \\ 
10.7099 & 0.3073 $\pm$ 0.0204 $\pm$ 0.0102 & 0.3305 $\pm$ 0.0219 $\pm$ 0.0110 & 1.2059 $\pm$ 0.0113 & 1.0758 $\pm$ 0.0008 \\ 
10.7164 & 0.3233 $\pm$ 0.0224 $\pm$ 0.0102 & 0.3478 $\pm$ 0.0241 $\pm$ 0.0110 & 1.1990 $\pm$ 0.0134 & 1.0758 $\pm$ 0.0008 \\ 
10.7189 & 0.3016 $\pm$ 0.0207 $\pm$ 0.0100 & 0.3245 $\pm$ 0.0223 $\pm$ 0.0108 & 1.2010 $\pm$ 0.0123 & 1.0758 $\pm$ 0.0008 \\ 
10.7249 & 0.3177 $\pm$ 0.0204 $\pm$ 0.0096 & 0.3418 $\pm$ 0.0219 $\pm$ 0.0103 & 1.2050 $\pm$ 0.0106 & 1.0759 $\pm$ 0.0008 \\ 
10.7294 & 0.3455 $\pm$ 0.0222 $\pm$ 0.0105 & 0.3718 $\pm$ 0.0239 $\pm$ 0.0113 & 1.2029 $\pm$ 0.0129 & 1.0759 $\pm$ 0.0008 \\ 
10.7322 & 0.3137 $\pm$ 0.0039 $\pm$ 0.0136 & 0.3375 $\pm$ 0.0042 $\pm$ 0.0147 & 1.1886 $\pm$ 0.0132 & 1.0760 $\pm$ 0.0008 \\ 
10.7344 & 0.3062 $\pm$ 0.0197 $\pm$ 0.0092 & 0.3295 $\pm$ 0.0212 $\pm$ 0.0099 & 1.1527 $\pm$ 0.0119 & 1.0761 $\pm$ 0.0008 \\ 
10.7409 & 0.2411 $\pm$ 0.0183 $\pm$ 0.0076 & 0.2594 $\pm$ 0.0196 $\pm$ 0.0082 & 1.0802 $\pm$ 0.0181 & 1.0760 $\pm$ 0.0008 \\ 
10.7449 & 0.1868 $\pm$ 0.0179 $\pm$ 0.0067 & 0.2010 $\pm$ 0.0192 $\pm$ 0.0072 & 1.0337 $\pm$ 0.0225 & 1.0760 $\pm$ 0.0008 \\ 
10.7459 & 0.1917 $\pm$ 0.0178 $\pm$ 0.0065 & 0.2062 $\pm$ 0.0191 $\pm$ 0.0070 & 1.0271 $\pm$ 0.0226 & 1.0759 $\pm$ 0.0008 \\ 
10.7535 & 0.2016 $\pm$ 0.0077 $\pm$ 0.0140 & 0.2169 $\pm$ 0.0082 $\pm$ 0.0151 & 1.0363 $\pm$ 0.0240 & 1.0757 $\pm$ 0.0008 \\ 
10.7539 & 0.2088 $\pm$ 0.0192 $\pm$ 0.0069 & 0.2246 $\pm$ 0.0207 $\pm$ 0.0074 & 1.0372 $\pm$ 0.0232 & 1.0757 $\pm$ 0.0008 \\ 
10.7579 & 0.1973 $\pm$ 0.0081 $\pm$ 0.0145 & 0.2122 $\pm$ 0.0087 $\pm$ 0.0156 & 1.0485 $\pm$ 0.0255 & 1.0756 $\pm$ 0.0008 \\ 
10.7589 & 0.1800 $\pm$ 0.0172 $\pm$ 0.0059 & 0.1935 $\pm$ 0.0185 $\pm$ 0.0063 & 1.0482 $\pm$ 0.0234 & 1.0755 $\pm$ 0.0008 \\ 
10.7637 & 0.1901 $\pm$ 0.0081 $\pm$ 0.0137 & 0.2045 $\pm$ 0.0087 $\pm$ 0.0147 & 1.0507 $\pm$ 0.0241 & 1.0754 $\pm$ 0.0008 \\ 
10.7639 & 0.1926 $\pm$ 0.0188 $\pm$ 0.0059 & 0.2071 $\pm$ 0.0202 $\pm$ 0.0064 & 1.0535 $\pm$ 0.0236 & 1.0754 $\pm$ 0.0008 \\ 
10.7677 & 0.2089 $\pm$ 0.0081 $\pm$ 0.0147 & 0.2247 $\pm$ 0.0087 $\pm$ 0.0158 & 1.0848 $\pm$ 0.0238 & 1.0753 $\pm$ 0.0008 \\ 
10.7679 & 0.2113 $\pm$ 0.0180 $\pm$ 0.0070 & 0.2272 $\pm$ 0.0194 $\pm$ 0.0075 & 1.0865 $\pm$ 0.0231 & 1.0753 $\pm$ 0.0008 \\ 
10.7711 & 0.2121 $\pm$ 0.0037 $\pm$ 0.0148 & 0.2280 $\pm$ 0.0039 $\pm$ 0.0159 & 1.1097 $\pm$ 0.0240 & 1.0752 $\pm$ 0.0008 \\ 
10.7716 & 0.2129 $\pm$ 0.0084 $\pm$ 0.0157 & 0.2289 $\pm$ 0.0091 $\pm$ 0.0169 & 1.1118 $\pm$ 0.0238 & 1.0752 $\pm$ 0.0008 \\ 
10.7739 & 0.2302 $\pm$ 0.0198 $\pm$ 0.0074 & 0.2475 $\pm$ 0.0213 $\pm$ 0.0080 & 1.1309 $\pm$ 0.0199 & 1.0752 $\pm$ 0.0008 \\ 
10.7760 & 0.2315 $\pm$ 0.0092 $\pm$ 0.0146 & 0.2489 $\pm$ 0.0099 $\pm$ 0.0157 & 1.1437 $\pm$ 0.0214 & 1.0751 $\pm$ 0.0008 \\ 
10.7789 & 0.2491 $\pm$ 0.0202 $\pm$ 0.0098 & 0.2679 $\pm$ 0.0217 $\pm$ 0.0106 & 1.1415 $\pm$ 0.0175 & 1.0751 $\pm$ 0.0008 \\ 
10.7820 & 0.2452 $\pm$ 0.0086 $\pm$ 0.0146 & 0.2636 $\pm$ 0.0092 $\pm$ 0.0157 & 1.1684 $\pm$ 0.0159 & 1.0750 $\pm$ 0.0008 \\ 
10.7839 & 0.2310 $\pm$ 0.0204 $\pm$ 0.0094 & 0.2483 $\pm$ 0.0219 $\pm$ 0.0101 & 1.1831 $\pm$ 0.0150 & 1.0750 $\pm$ 0.0008 \\ 
10.7871 & 0.2979 $\pm$ 0.0087 $\pm$ 0.0160 & 0.3202 $\pm$ 0.0093 $\pm$ 0.0172 & 1.1999 $\pm$ 0.0144 & 1.0750 $\pm$ 0.0008 \\ 
10.7889 & 0.2861 $\pm$ 0.0209 $\pm$ 0.0111 & 0.3076 $\pm$ 0.0225 $\pm$ 0.0119 & 1.2116 $\pm$ 0.0139 & 1.0750 $\pm$ 0.0008 \\ 
10.7920 & 0.3106 $\pm$ 0.0089 $\pm$ 0.0157 & 0.3339 $\pm$ 0.0096 $\pm$ 0.0169 & 1.2251 $\pm$ 0.0130 & 1.0750 $\pm$ 0.0008 \\ 
10.7949 & 0.2869 $\pm$ 0.0213 $\pm$ 0.0102 & 0.3084 $\pm$ 0.0229 $\pm$ 0.0109 & 1.2296 $\pm$ 0.0127 & 1.0750 $\pm$ 0.0008 \\ 
10.7955 & 0.3145 $\pm$ 0.0083 $\pm$ 0.0153 & 0.3381 $\pm$ 0.0089 $\pm$ 0.0164 & 1.2311 $\pm$ 0.0130 & 1.0750 $\pm$ 0.0008 \\ 
10.7979 & 0.2978 $\pm$ 0.0218 $\pm$ 0.0112 & 0.3201 $\pm$ 0.0234 $\pm$ 0.0120 & 1.2407 $\pm$ 0.0132 & 1.0750 $\pm$ 0.0008 \\ 
10.7999 & 0.3540 $\pm$ 0.0082 $\pm$ 0.0156 & 0.3805 $\pm$ 0.0088 $\pm$ 0.0167 & 1.2498 $\pm$ 0.0106 & 1.0750 $\pm$ 0.0008 \\ 
10.8063 & 0.3655 $\pm$ 0.0096 $\pm$ 0.0164 & 0.3930 $\pm$ 0.0104 $\pm$ 0.0176 & 1.2668 $\pm$ 0.0111 & 1.0750 $\pm$ 0.0008 \\ 
10.8075 & 0.3675 $\pm$ 0.0234 $\pm$ 0.0126 & 0.3951 $\pm$ 0.0251 $\pm$ 0.0135 & 1.2665 $\pm$ 0.0102 & 1.0751 $\pm$ 0.0008 \\ 
10.8107 & 0.3486 $\pm$ 0.0093 $\pm$ 0.0155 & 0.3748 $\pm$ 0.0100 $\pm$ 0.0166 & 1.2616 $\pm$ 0.0096 & 1.0751 $\pm$ 0.0008 \\ 
10.8130 & 0.3876 $\pm$ 0.0222 $\pm$ 0.0127 & 0.4167 $\pm$ 0.0238 $\pm$ 0.0136 & 1.2522 $\pm$ 0.0082 & 1.0751 $\pm$ 0.0008 \\ 
10.8157 & 0.3846 $\pm$ 0.0098 $\pm$ 0.0167 & 0.4136 $\pm$ 0.0105 $\pm$ 0.0180 & 1.2478 $\pm$ 0.0108 & 1.0752 $\pm$ 0.0008 \\ 
10.8175 & 0.3389 $\pm$ 0.0215 $\pm$ 0.0109 & 0.3644 $\pm$ 0.0231 $\pm$ 0.0118 & 1.2436 $\pm$ 0.0106 & 1.0752 $\pm$ 0.0008 \\ 
10.8210 & 0.3694 $\pm$ 0.0094 $\pm$ 0.0167 & 0.3972 $\pm$ 0.0101 $\pm$ 0.0180 & 1.2387 $\pm$ 0.0102 & 1.0751 $\pm$ 0.0008 \\ 
10.8220 & 0.3487 $\pm$ 0.0216 $\pm$ 0.0120 & 0.3749 $\pm$ 0.0232 $\pm$ 0.0129 & 1.2387 $\pm$ 0.0105 & 1.0751 $\pm$ 0.0008 \\ 
10.8259 & 0.3796 $\pm$ 0.0094 $\pm$ 0.0160 & 0.4081 $\pm$ 0.0101 $\pm$ 0.0172 & 1.2536 $\pm$ 0.0096 & 1.0751 $\pm$ 0.0008 \\ 
10.8275 & 0.3862 $\pm$ 0.0230 $\pm$ 0.0142 & 0.4152 $\pm$ 0.0248 $\pm$ 0.0153 & 1.2607 $\pm$ 0.0094 & 1.0750 $\pm$ 0.0008 \\ 
10.8304 & 0.4252 $\pm$ 0.0090 $\pm$ 0.0167 & 0.4571 $\pm$ 0.0097 $\pm$ 0.0180 & 1.2721 $\pm$ 0.0102 & 1.0750 $\pm$ 0.0008 \\ 
10.8320 & 0.3762 $\pm$ 0.0215 $\pm$ 0.0147 & 0.4044 $\pm$ 0.0231 $\pm$ 0.0158 & 1.2773 $\pm$ 0.0090 & 1.0750 $\pm$ 0.0008 \\ 
10.8332 & 0.4486 $\pm$ 0.0090 $\pm$ 0.0161 & 0.4823 $\pm$ 0.0097 $\pm$ 0.0173 & 1.2859 $\pm$ 0.0078 & 1.0750 $\pm$ 0.0008 \\ 
10.8375 & 0.4426 $\pm$ 0.0235 $\pm$ 0.0161 & 0.4759 $\pm$ 0.0252 $\pm$ 0.0174 & 1.3047 $\pm$ 0.0088 & 1.0751 $\pm$ 0.0008 \\ 
10.8396 & 0.4691 $\pm$ 0.0102 $\pm$ 0.0167 & 0.5044 $\pm$ 0.0109 $\pm$ 0.0179 & 1.3079 $\pm$ 0.0087 & 1.0751 $\pm$ 0.0008 \\ 
10.8415 & 0.5296 $\pm$ 0.0251 $\pm$ 0.0203 & 0.5695 $\pm$ 0.0270 $\pm$ 0.0218 & 1.3094 $\pm$ 0.0085 & 1.0752 $\pm$ 0.0008 \\ 
10.8450 & 0.4959 $\pm$ 0.0097 $\pm$ 0.0198 & 0.5332 $\pm$ 0.0104 $\pm$ 0.0212 & 1.3056 $\pm$ 0.0074 & 1.0753 $\pm$ 0.0008 \\ 
10.8455 & 0.4758 $\pm$ 0.0232 $\pm$ 0.0159 & 0.5117 $\pm$ 0.0250 $\pm$ 0.0171 & 1.3049 $\pm$ 0.0076 & 1.0753 $\pm$ 0.0008 \\ 
10.8494 & 0.5080 $\pm$ 0.0093 $\pm$ 0.0184 & 0.5464 $\pm$ 0.0100 $\pm$ 0.0198 & 1.2981 $\pm$ 0.0114 & 1.0755 $\pm$ 0.0008 \\ 
10.8497 & 0.4978 $\pm$ 0.0030 $\pm$ 0.0180 & 0.5353 $\pm$ 0.0032 $\pm$ 0.0194 & 1.2978 $\pm$ 0.0109 & 1.0755 $\pm$ 0.0008 \\ 
10.8528 & 0.5113 $\pm$ 0.0099 $\pm$ 0.0210 & 0.5499 $\pm$ 0.0106 $\pm$ 0.0226 & 1.2968 $\pm$ 0.0121 & 1.0755 $\pm$ 0.0008 \\ 
10.8577 & 0.5126 $\pm$ 0.0099 $\pm$ 0.0173 & 0.5514 $\pm$ 0.0106 $\pm$ 0.0186 & 1.2940 $\pm$ 0.0075 & 1.0757 $\pm$ 0.0008 \\ 
10.8589 & 0.5186 $\pm$ 0.0035 $\pm$ 0.0157 & 0.5579 $\pm$ 0.0037 $\pm$ 0.0168 & 1.2955 $\pm$ 0.0074 & 1.0757 $\pm$ 0.0008 \\ 
10.8600 & 0.5043 $\pm$ 0.0232 $\pm$ 0.0176 & 0.5425 $\pm$ 0.0250 $\pm$ 0.0189 & 1.2960 $\pm$ 0.0077 & 1.0757 $\pm$ 0.0008 \\ 
10.8605 & 0.5648 $\pm$ 0.0248 $\pm$ 0.0195 & 0.6076 $\pm$ 0.0266 $\pm$ 0.0209 & 1.2966 $\pm$ 0.0080 & 1.0757 $\pm$ 0.0008 \\ 
10.8639 & 0.5314 $\pm$ 0.0107 $\pm$ 0.0167 & 0.5718 $\pm$ 0.0115 $\pm$ 0.0180 & 1.3026 $\pm$ 0.0088 & 1.0759 $\pm$ 0.0008 \\ 
10.8645 & 0.5586 $\pm$ 0.0263 $\pm$ 0.0182 & 0.6010 $\pm$ 0.0283 $\pm$ 0.0196 & 1.3018 $\pm$ 0.0097 & 1.0759 $\pm$ 0.0008 \\ 
10.8667 & 0.5538 $\pm$ 0.0099 $\pm$ 0.0177 & 0.5958 $\pm$ 0.0106 $\pm$ 0.0191 & 1.2952 $\pm$ 0.0065 & 1.0760 $\pm$ 0.0008 \\ 
10.8690 & 0.5474 $\pm$ 0.0033 $\pm$ 0.0153 & 0.5890 $\pm$ 0.0035 $\pm$ 0.0164 & 1.2914 $\pm$ 0.0076 & 1.0761 $\pm$ 0.0008 \\ 
10.8695 & 0.5462 $\pm$ 0.0032 $\pm$ 0.0162 & 0.5878 $\pm$ 0.0034 $\pm$ 0.0174 & 1.2909 $\pm$ 0.0064 & 1.0761 $\pm$ 0.0008 \\ 
10.8700 & 0.5511 $\pm$ 0.0249 $\pm$ 0.0176 & 0.5931 $\pm$ 0.0268 $\pm$ 0.0190 & 1.2902 $\pm$ 0.0069 & 1.0761 $\pm$ 0.0008 \\ 
10.8752 & 0.5472 $\pm$ 0.0071 $\pm$ 0.0165 & 0.5891 $\pm$ 0.0076 $\pm$ 0.0177 & 1.2829 $\pm$ 0.0059 & 1.0764 $\pm$ 0.0008 \\ 
10.8760 & 0.5630 $\pm$ 0.0230 $\pm$ 0.0179 & 0.6060 $\pm$ 0.0248 $\pm$ 0.0192 & 1.2822 $\pm$ 0.0057 & 1.0765 $\pm$ 0.0008 \\ 
10.8785 & 0.5436 $\pm$ 0.0036 $\pm$ 0.0157 & 0.5853 $\pm$ 0.0039 $\pm$ 0.0169 & 1.2767 $\pm$ 0.0071 & 1.0767 $\pm$ 0.0008 \\ 
10.8788 & 0.5538 $\pm$ 0.0062 $\pm$ 0.0159 & 0.5962 $\pm$ 0.0067 $\pm$ 0.0171 & 1.2748 $\pm$ 0.0073 & 1.0767 $\pm$ 0.0008 \\ 
10.8810 & 0.5442 $\pm$ 0.0212 $\pm$ 0.0163 & 0.5860 $\pm$ 0.0228 $\pm$ 0.0176 & 1.2441 $\pm$ 0.0054 & 1.0768 $\pm$ 0.0008 \\ 
10.8860 & 0.4648 $\pm$ 0.0086 $\pm$ 0.0148 & 0.5006 $\pm$ 0.0092 $\pm$ 0.0160 & 1.1993 $\pm$ 0.0073 & 1.0770 $\pm$ 0.0008 \\ 
10.8880 & 0.4162 $\pm$ 0.0206 $\pm$ 0.0127 & 0.4483 $\pm$ 0.0222 $\pm$ 0.0137 & 1.1740 $\pm$ 0.0085 & 1.0771 $\pm$ 0.0008 \\ 
10.8889 & 0.4254 $\pm$ 0.0034 $\pm$ 0.0141 & 0.4582 $\pm$ 0.0037 $\pm$ 0.0152 & 1.1625 $\pm$ 0.0084 & 1.0771 $\pm$ 0.0008 \\ 
10.8918 & 0.4009 $\pm$ 0.0087 $\pm$ 0.0150 & 0.4319 $\pm$ 0.0094 $\pm$ 0.0161 & 1.1326 $\pm$ 0.0093 & 1.0771 $\pm$ 0.0008 \\ 
10.8940 & 0.3405 $\pm$ 0.0179 $\pm$ 0.0103 & 0.3667 $\pm$ 0.0193 $\pm$ 0.0111 & 1.1132 $\pm$ 0.0103 & 1.0771 $\pm$ 0.0008 \\ 
10.8962 & 0.3514 $\pm$ 0.0088 $\pm$ 0.0174 & 0.3784 $\pm$ 0.0095 $\pm$ 0.0187 & 1.0909 $\pm$ 0.0135 & 1.0770 $\pm$ 0.0008 \\ 
10.8985 & 0.3207 $\pm$ 0.0029 $\pm$ 0.0141 & 0.3453 $\pm$ 0.0031 $\pm$ 0.0152 & 1.0754 $\pm$ 0.0120 & 1.0770 $\pm$ 0.0008 \\ 
10.9009 & 0.3038 $\pm$ 0.0084 $\pm$ 0.0139 & 0.3272 $\pm$ 0.0090 $\pm$ 0.0150 & 1.0637 $\pm$ 0.0130 & 1.0769 $\pm$ 0.0008 \\ 
10.9055 & 0.2906 $\pm$ 0.0184 $\pm$ 0.0103 & 0.3129 $\pm$ 0.0198 $\pm$ 0.0111 & 1.0526 $\pm$ 0.0124 & 1.0768 $\pm$ 0.0008 \\ 
10.9056 & 0.2850 $\pm$ 0.0086 $\pm$ 0.0122 & 0.3069 $\pm$ 0.0093 $\pm$ 0.0132 & 1.0525 $\pm$ 0.0126 & 1.0768 $\pm$ 0.0008 \\ 
10.9077 & 0.2742 $\pm$ 0.0041 $\pm$ 0.0108 & 0.2952 $\pm$ 0.0044 $\pm$ 0.0116 & 1.0572 $\pm$ 0.0149 & 1.0768 $\pm$ 0.0008 \\ 
10.9095 & 0.3449 $\pm$ 0.0161 $\pm$ 0.0128 & 0.3713 $\pm$ 0.0173 $\pm$ 0.0137 & 1.0516 $\pm$ 0.0176 & 1.0768 $\pm$ 0.0008 \\ 
10.9104 & 0.2689 $\pm$ 0.0087 $\pm$ 0.0115 & 0.2896 $\pm$ 0.0094 $\pm$ 0.0124 & 1.0502 $\pm$ 0.0191 & 1.0768 $\pm$ 0.0008 \\ 
10.9110 & 0.2589 $\pm$ 0.0356 $\pm$ 0.0101 & 0.2787 $\pm$ 0.0383 $\pm$ 0.0108 & 1.0495 $\pm$ 0.0207 & 1.0768 $\pm$ 0.0008 \\ 
10.9135 & 0.2626 $\pm$ 0.0193 $\pm$ 0.0101 & 0.2827 $\pm$ 0.0208 $\pm$ 0.0108 & 1.0178 $\pm$ 0.0198 & 1.0767 $\pm$ 0.0008 \\ 
10.9152 & 0.2526 $\pm$ 0.0089 $\pm$ 0.0120 & 0.2720 $\pm$ 0.0096 $\pm$ 0.0129 & 0.9982 $\pm$ 0.0208 & 1.0767 $\pm$ 0.0008 \\ 
10.9185 & 0.2201 $\pm$ 0.0185 $\pm$ 0.0087 & 0.2370 $\pm$ 0.0199 $\pm$ 0.0094 & 0.9954 $\pm$ 0.0172 & 1.0766 $\pm$ 0.0008 \\ 
10.9215 & 0.2373 $\pm$ 0.0084 $\pm$ 0.0126 & 0.2555 $\pm$ 0.0091 $\pm$ 0.0136 & 0.9966 $\pm$ 0.0193 & 1.0765 $\pm$ 0.0008 \\ 
10.9235 & 0.2281 $\pm$ 0.0182 $\pm$ 0.0084 & 0.2455 $\pm$ 0.0196 $\pm$ 0.0091 & 1.0018 $\pm$ 0.0159 & 1.0764 $\pm$ 0.0008 \\ 
10.9250 & 0.2410 $\pm$ 0.0079 $\pm$ 0.0124 & 0.2594 $\pm$ 0.0085 $\pm$ 0.0134 & 1.0058 $\pm$ 0.0151 & 1.0764 $\pm$ 0.0008 \\ 
10.9313 & 0.2338 $\pm$ 0.0081 $\pm$ 0.0126 & 0.2517 $\pm$ 0.0088 $\pm$ 0.0136 & 1.0224 $\pm$ 0.0164 & 1.0763 $\pm$ 0.0008 \\ 
10.9315 & 0.2345 $\pm$ 0.0168 $\pm$ 0.0086 & 0.2524 $\pm$ 0.0181 $\pm$ 0.0093 & 1.0221 $\pm$ 0.0159 & 1.0763 $\pm$ 0.0008 \\ 
10.9348 & 0.2344 $\pm$ 0.0080 $\pm$ 0.0123 & 0.2523 $\pm$ 0.0086 $\pm$ 0.0132 & 1.0218 $\pm$ 0.0162 & 1.0763 $\pm$ 0.0008 \\ 
10.9365 & 0.2375 $\pm$ 0.0182 $\pm$ 0.0089 & 0.2556 $\pm$ 0.0196 $\pm$ 0.0095 & 1.0239 $\pm$ 0.0147 & 1.0762 $\pm$ 0.0008 \\ 
10.9386 & 0.2214 $\pm$ 0.0173 $\pm$ 0.0089 & 0.2383 $\pm$ 0.0186 $\pm$ 0.0095 & 1.0245 $\pm$ 0.0152 & 1.0762 $\pm$ 0.0008 \\ 
10.9400 & 0.2386 $\pm$ 0.0084 $\pm$ 0.0124 & 0.2568 $\pm$ 0.0090 $\pm$ 0.0133 & 1.0242 $\pm$ 0.0171 & 1.0762 $\pm$ 0.0008 \\ 
10.9444 & 0.2329 $\pm$ 0.0082 $\pm$ 0.0129 & 0.2507 $\pm$ 0.0088 $\pm$ 0.0138 & 1.0208 $\pm$ 0.0171 & 1.0761 $\pm$ 0.0008 \\ 
10.9456 & 0.2177 $\pm$ 0.0179 $\pm$ 0.0084 & 0.2343 $\pm$ 0.0193 $\pm$ 0.0090 & 1.0214 $\pm$ 0.0181 & 1.0761 $\pm$ 0.0008 \\ 
10.9493 & 0.2346 $\pm$ 0.0080 $\pm$ 0.0119 & 0.2525 $\pm$ 0.0086 $\pm$ 0.0128 & 1.0276 $\pm$ 0.0166 & 1.0761 $\pm$ 0.0008 \\ 
10.9501 & 0.2125 $\pm$ 0.0173 $\pm$ 0.0080 & 0.2286 $\pm$ 0.0186 $\pm$ 0.0086 & 1.0294 $\pm$ 0.0158 & 1.0760 $\pm$ 0.0008 \\ 
10.9536 & 0.2382 $\pm$ 0.0189 $\pm$ 0.0090 & 0.2563 $\pm$ 0.0203 $\pm$ 0.0096 & 1.0354 $\pm$ 0.0172 & 1.0760 $\pm$ 0.0008 \\ 
10.9537 & 0.2328 $\pm$ 0.0083 $\pm$ 0.0118 & 0.2505 $\pm$ 0.0089 $\pm$ 0.0127 & 1.0358 $\pm$ 0.0173 & 1.0760 $\pm$ 0.0008 \\ 
10.9576 & 0.2163 $\pm$ 0.0176 $\pm$ 0.0077 & 0.2327 $\pm$ 0.0190 $\pm$ 0.0083 & 1.0388 $\pm$ 0.0154 & 1.0759 $\pm$ 0.0008 \\ 
10.9590 & 0.2377 $\pm$ 0.0080 $\pm$ 0.0138 & 0.2557 $\pm$ 0.0086 $\pm$ 0.0149 & 1.0356 $\pm$ 0.0149 & 1.0759 $\pm$ 0.0008 \\ 
10.9633 & 0.2274 $\pm$ 0.0078 $\pm$ 0.0126 & 0.2447 $\pm$ 0.0084 $\pm$ 0.0136 & 1.0456 $\pm$ 0.0139 & 1.0758 $\pm$ 0.0008 \\ 
10.9641 & 0.2206 $\pm$ 0.0085 $\pm$ 0.0086 & 0.2373 $\pm$ 0.0091 $\pm$ 0.0092 & 1.0495 $\pm$ 0.0136 & 1.0758 $\pm$ 0.0008 \\ 
10.9674 & 0.2236 $\pm$ 0.0078 $\pm$ 0.0116 & 0.2405 $\pm$ 0.0084 $\pm$ 0.0125 & 1.0459 $\pm$ 0.0154 & 1.0758 $\pm$ 0.0008 \\ 
10.9686 & 0.2613 $\pm$ 0.0196 $\pm$ 0.0091 & 0.2811 $\pm$ 0.0211 $\pm$ 0.0098 & 1.0443 $\pm$ 0.0166 & 1.0757 $\pm$ 0.0008 \\ 
10.9726 & 0.2036 $\pm$ 0.0174 $\pm$ 0.0072 & 0.2190 $\pm$ 0.0187 $\pm$ 0.0078 & 1.0578 $\pm$ 0.0143 & 1.0756 $\pm$ 0.0008 \\ 
10.9727 & 0.2345 $\pm$ 0.0078 $\pm$ 0.0123 & 0.2522 $\pm$ 0.0084 $\pm$ 0.0132 & 1.0579 $\pm$ 0.0139 & 1.0756 $\pm$ 0.0008 \\ 
10.9773 & 0.2441 $\pm$ 0.0078 $\pm$ 0.0123 & 0.2625 $\pm$ 0.0084 $\pm$ 0.0132 & 1.0750 $\pm$ 0.0114 & 1.0755 $\pm$ 0.0008 \\ 
10.9775 & 0.2444 $\pm$ 0.0028 $\pm$ 0.0113 & 0.2629 $\pm$ 0.0030 $\pm$ 0.0122 & 1.0784 $\pm$ 0.0113 & 1.0755 $\pm$ 0.0008 \\ 
10.9791 & 0.2469 $\pm$ 0.0087 $\pm$ 0.0086 & 0.2655 $\pm$ 0.0093 $\pm$ 0.0093 & 1.1028 $\pm$ 0.0116 & 1.0754 $\pm$ 0.0008 \\ 
10.9833 & 0.2681 $\pm$ 0.0084 $\pm$ 0.0123 & 0.2883 $\pm$ 0.0090 $\pm$ 0.0132 & 1.1385 $\pm$ 0.0123 & 1.0753 $\pm$ 0.0008 \\ 
10.9836 & 0.3097 $\pm$ 0.0208 $\pm$ 0.0106 & 0.3331 $\pm$ 0.0223 $\pm$ 0.0114 & 1.1408 $\pm$ 0.0126 & 1.0753 $\pm$ 0.0008 \\ 
10.9873 & 0.2927 $\pm$ 0.0081 $\pm$ 0.0133 & 0.3148 $\pm$ 0.0087 $\pm$ 0.0144 & 1.1747 $\pm$ 0.0097 & 1.0753 $\pm$ 0.0008 \\ 
10.9901 & 0.3184 $\pm$ 0.0206 $\pm$ 0.0104 & 0.3424 $\pm$ 0.0221 $\pm$ 0.0112 & 1.1936 $\pm$ 0.0101 & 1.0752 $\pm$ 0.0008 \\ 
10.9919 & 0.3344 $\pm$ 0.0034 $\pm$ 0.0121 & 0.3595 $\pm$ 0.0037 $\pm$ 0.0130 & 1.2094 $\pm$ 0.0101 & 1.0752 $\pm$ 0.0008 \\ 
10.9927 & 0.3523 $\pm$ 0.0084 $\pm$ 0.0141 & 0.3788 $\pm$ 0.0090 $\pm$ 0.0152 & 1.2200 $\pm$ 0.0082 & 1.0752 $\pm$ 0.0008 \\ 
10.9936 & 0.3883 $\pm$ 0.0195 $\pm$ 0.0124 & 0.4175 $\pm$ 0.0209 $\pm$ 0.0133 & 1.2308 $\pm$ 0.0068 & 1.0752 $\pm$ 0.0008 \\ 
10.9975 & 0.3853 $\pm$ 0.0089 $\pm$ 0.0139 & 0.4143 $\pm$ 0.0095 $\pm$ 0.0149 & 1.2700 $\pm$ 0.0063 & 1.0752 $\pm$ 0.0008 \\ 
10.9991 & 0.4450 $\pm$ 0.0101 $\pm$ 0.0143 & 0.4785 $\pm$ 0.0108 $\pm$ 0.0154 & 1.2785 $\pm$ 0.0068 & 1.0753 $\pm$ 0.0008 \\ 
11.0013 & 0.4512 $\pm$ 0.0092 $\pm$ 0.0136 & 0.4852 $\pm$ 0.0099 $\pm$ 0.0147 & 1.2880 $\pm$ 0.0053 & 1.0754 $\pm$ 0.0008 \\ 
11.0041 & 0.5033 $\pm$ 0.0109 $\pm$ 0.0152 & 0.5413 $\pm$ 0.0117 $\pm$ 0.0164 & 1.2914 $\pm$ 0.0043 & 1.0755 $\pm$ 0.0008 \\ 
11.0068 & 0.4808 $\pm$ 0.0028 $\pm$ 0.0135 & 0.5171 $\pm$ 0.0030 $\pm$ 0.0145 & 1.2940 $\pm$ 0.0048 & 1.0756 $\pm$ 0.0008 \\ 
11.0069 & 0.4848 $\pm$ 0.0089 $\pm$ 0.0144 & 0.5215 $\pm$ 0.0096 $\pm$ 0.0154 & 1.2935 $\pm$ 0.0048 & 1.0757 $\pm$ 0.0008 \\ 
11.0086 & 0.4884 $\pm$ 0.0097 $\pm$ 0.0138 & 0.5254 $\pm$ 0.0104 $\pm$ 0.0148 & 1.2901 $\pm$ 0.0051 & 1.0758 $\pm$ 0.0008 \\ 
11.0121 & 0.4962 $\pm$ 0.0098 $\pm$ 0.0143 & 0.5339 $\pm$ 0.0105 $\pm$ 0.0154 & 1.2823 $\pm$ 0.0067 & 1.0759 $\pm$ 0.0008 \\ 
11.0151 & 0.5246 $\pm$ 0.0236 $\pm$ 0.0149 & 0.5645 $\pm$ 0.0254 $\pm$ 0.0160 & 1.2729 $\pm$ 0.0063 & 1.0761 $\pm$ 0.0008 \\ 
11.0164 & 0.4900 $\pm$ 0.0033 $\pm$ 0.0130 & 0.5274 $\pm$ 0.0036 $\pm$ 0.0140 & 1.2659 $\pm$ 0.0059 & 1.0762 $\pm$ 0.0008 \\ 
11.0188 & 0.4662 $\pm$ 0.0093 $\pm$ 0.0147 & 0.5018 $\pm$ 0.0100 $\pm$ 0.0159 & 1.2456 $\pm$ 0.0047 & 1.0763 $\pm$ 0.0008 \\ 
11.0191 & 0.4761 $\pm$ 0.0094 $\pm$ 0.0131 & 0.5125 $\pm$ 0.0101 $\pm$ 0.0141 & 1.2423 $\pm$ 0.0049 & 1.0763 $\pm$ 0.0008 \\ 
11.0214 & 0.4621 $\pm$ 0.0095 $\pm$ 0.0137 & 0.4974 $\pm$ 0.0102 $\pm$ 0.0147 & 1.2195 $\pm$ 0.0061 & 1.0764 $\pm$ 0.0008 \\ 
11.0220 & 0.4508 $\pm$ 0.0030 $\pm$ 0.0124 & 0.4852 $\pm$ 0.0032 $\pm$ 0.0133 & 1.2184 $\pm$ 0.0058 & 1.0764 $\pm$ 0.0008 \\ 
11.0241 & 0.4253 $\pm$ 0.0096 $\pm$ 0.0136 & 0.4579 $\pm$ 0.0103 $\pm$ 0.0146 & 1.2098 $\pm$ 0.0086 & 1.0764 $\pm$ 0.0008 \\ 
11.0266 & 0.4254 $\pm$ 0.0215 $\pm$ 0.0133 & 0.4579 $\pm$ 0.0231 $\pm$ 0.0143 & 1.1998 $\pm$ 0.0084 & 1.0764 $\pm$ 0.0008 \\ 
11.0269 & 0.4425 $\pm$ 0.0093 $\pm$ 0.0134 & 0.4763 $\pm$ 0.0100 $\pm$ 0.0144 & 1.1998 $\pm$ 0.0077 & 1.0764 $\pm$ 0.0008 \\ 
11.0313 & 0.4161 $\pm$ 0.0101 $\pm$ 0.0133 & 0.4479 $\pm$ 0.0109 $\pm$ 0.0143 & 1.1912 $\pm$ 0.0120 & 1.0765 $\pm$ 0.0008 \\ 
11.0331 & 0.4199 $\pm$ 0.0200 $\pm$ 0.0128 & 0.4520 $\pm$ 0.0215 $\pm$ 0.0138 & 1.1813 $\pm$ 0.0064 & 1.0765 $\pm$ 0.0008 \\ 
11.0386 & 0.3870 $\pm$ 0.0089 $\pm$ 0.0132 & 0.4166 $\pm$ 0.0096 $\pm$ 0.0142 & 1.1593 $\pm$ 0.0087 & 1.0766 $\pm$ 0.0008 \\ 
11.0401 & 0.3947 $\pm$ 0.0198 $\pm$ 0.0131 & 0.4250 $\pm$ 0.0213 $\pm$ 0.0141 & 1.1589 $\pm$ 0.0078 & 1.0766 $\pm$ 0.0008 \\ 
11.0402 & 0.3815 $\pm$ 0.0087 $\pm$ 0.0132 & 0.4107 $\pm$ 0.0093 $\pm$ 0.0143 & 1.1584 $\pm$ 0.0079 & 1.0766 $\pm$ 0.0008 \\ 
11.0446 & 0.3528 $\pm$ 0.0198 $\pm$ 0.0123 & 0.3799 $\pm$ 0.0213 $\pm$ 0.0132 & 1.1490 $\pm$ 0.0102 & 1.0766 $\pm$ 0.0008 \\ 
11.0474 & 0.3720 $\pm$ 0.0090 $\pm$ 0.0111 & 0.4004 $\pm$ 0.0097 $\pm$ 0.0120 & 1.1381 $\pm$ 0.0127 & 1.0766 $\pm$ 0.0008 \\ 
11.0491 & 0.4037 $\pm$ 0.0215 $\pm$ 0.0146 & 0.4346 $\pm$ 0.0231 $\pm$ 0.0157 & 1.1355 $\pm$ 0.0140 & 1.0766 $\pm$ 0.0008 \\ 
11.0531 & 0.3185 $\pm$ 0.0200 $\pm$ 0.0124 & 0.3429 $\pm$ 0.0216 $\pm$ 0.0133 & 1.1315 $\pm$ 0.0160 & 1.0765 $\pm$ 0.0008 \\ 
11.0581 & 0.3572 $\pm$ 0.0194 $\pm$ 0.0139 & 0.3845 $\pm$ 0.0209 $\pm$ 0.0149 & 1.1228 $\pm$ 0.0149 & 1.0765 $\pm$ 0.0008 \\ 
11.0636 & 0.3374 $\pm$ 0.0188 $\pm$ 0.0139 & 0.3632 $\pm$ 0.0202 $\pm$ 0.0149 & 1.1193 $\pm$ 0.0163 & 1.0765 $\pm$ 0.0008 \\ 
11.0686 & 0.3265 $\pm$ 0.0196 $\pm$ 0.0136 & 0.3515 $\pm$ 0.0211 $\pm$ 0.0146 & 1.1182 $\pm$ 0.0154 & 1.0765 $\pm$ 0.0008 \\ 
11.0721 & 0.3545 $\pm$ 0.0198 $\pm$ 0.0136 & 0.3816 $\pm$ 0.0213 $\pm$ 0.0147 & 1.1179 $\pm$ 0.0142 & 1.0765 $\pm$ 0.0008 \\ 
11.0787 & 0.3238 $\pm$ 0.0182 $\pm$ 0.0132 & 0.3486 $\pm$ 0.0195 $\pm$ 0.0142 & 1.1225 $\pm$ 0.0139 & 1.0764 $\pm$ 0.0008 \\ 
11.0827 & 0.3324 $\pm$ 0.0166 $\pm$ 0.0131 & 0.3578 $\pm$ 0.0178 $\pm$ 0.0141 & 1.1262 $\pm$ 0.0142 & 1.0764 $\pm$ 0.0008 \\ 
11.0877 & 0.3466 $\pm$ 0.0201 $\pm$ 0.0131 & 0.3731 $\pm$ 0.0216 $\pm$ 0.0141 & 1.1254 $\pm$ 0.0135 & 1.0764 $\pm$ 0.0008 \\ 
11.0932 & 0.3593 $\pm$ 0.0201 $\pm$ 0.0130 & 0.3868 $\pm$ 0.0217 $\pm$ 0.0140 & 1.1296 $\pm$ 0.0143 & 1.0764 $\pm$ 0.0008 \\ 
11.0982 & 0.3286 $\pm$ 0.0196 $\pm$ 0.0127 & 0.3537 $\pm$ 0.0211 $\pm$ 0.0137 & 1.1245 $\pm$ 0.0127 & 1.0765 $\pm$ 0.0008 \\ 
11.1012 & 0.3368 $\pm$ 0.0191 $\pm$ 0.0127 & 0.3626 $\pm$ 0.0206 $\pm$ 0.0137 & 1.1186 $\pm$ 0.0121 & 1.0765 $\pm$ 0.0008 \\ 
11.1077 & 0.3366 $\pm$ 0.0090 $\pm$ 0.0126 & 0.3623 $\pm$ 0.0097 $\pm$ 0.0136 & 1.1144 $\pm$ 0.0125 & 1.0764 $\pm$ 0.0008 \\ 
11.1092 & 0.2959 $\pm$ 0.0193 $\pm$ 0.0113 & 0.3186 $\pm$ 0.0208 $\pm$ 0.0121 & 1.1151 $\pm$ 0.0131 & 1.0764 $\pm$ 0.0008 \\ 
11.1112 & 0.3445 $\pm$ 0.0188 $\pm$ 0.0126 & 0.3709 $\pm$ 0.0202 $\pm$ 0.0136 & 1.1190 $\pm$ 0.0134 & 1.0764 $\pm$ 0.0008 \\ 
11.1162 & 0.3463 $\pm$ 0.0188 $\pm$ 0.0126 & 0.3728 $\pm$ 0.0202 $\pm$ 0.0135 & 1.1289 $\pm$ 0.0171 & 1.0764 $\pm$ 0.0008 \\ 
11.1217 & 0.3505 $\pm$ 0.0210 $\pm$ 0.0126 & 0.3773 $\pm$ 0.0227 $\pm$ 0.0135 & 1.1412 $\pm$ 0.0233 & 1.0764 $\pm$ 0.0008 \\ 
11.1262 & 0.3547 $\pm$ 0.0564 $\pm$ 0.0125 & 0.3819 $\pm$ 0.0608 $\pm$ 0.0134 & 1.1292 $\pm$ 0.0192 & 1.0765 $\pm$ 0.0008 \\ 
11.1277 & 0.3472 $\pm$ 0.0202 $\pm$ 0.0125 & 0.3737 $\pm$ 0.0217 $\pm$ 0.0134 & 1.1266 $\pm$ 0.0178 & 1.0765 $\pm$ 0.0008 \\ 
11.1327 & 0.3315 $\pm$ 0.0190 $\pm$ 0.0124 & 0.3568 $\pm$ 0.0205 $\pm$ 0.0134 & 1.1266 $\pm$ 0.0150 & 1.0765 $\pm$ 0.0008 \\ 
11.1347 & 0.3343 $\pm$ 0.0195 $\pm$ 0.0124 & 0.3599 $\pm$ 0.0210 $\pm$ 0.0134 & 1.1303 $\pm$ 0.0124 & 1.0765 $\pm$ 0.0008 \\ 
11.1432 & 0.3591 $\pm$ 0.0203 $\pm$ 0.0136 & 0.3865 $\pm$ 0.0219 $\pm$ 0.0146 & 1.1286 $\pm$ 0.0151 & 1.0765 $\pm$ 0.0008 \\ 
11.1467 & 0.3549 $\pm$ 0.0199 $\pm$ 0.0124 & 0.3820 $\pm$ 0.0214 $\pm$ 0.0134 & 1.1337 $\pm$ 0.0126 & 1.0765 $\pm$ 0.0008 \\ 
11.1497 & 0.3403 $\pm$ 0.0187 $\pm$ 0.0124 & 0.3663 $\pm$ 0.0201 $\pm$ 0.0133 & 1.1367 $\pm$ 0.0137 & 1.0765 $\pm$ 0.0008 \\ 
11.1552 & 0.3434 $\pm$ 0.0196 $\pm$ 0.0123 & 0.3696 $\pm$ 0.0211 $\pm$ 0.0132 & 1.1255 $\pm$ 0.0137 & 1.0765 $\pm$ 0.0008 \\ 
11.1582 & 0.3777 $\pm$ 0.0280 $\pm$ 0.0135 & 0.4067 $\pm$ 0.0301 $\pm$ 0.0146 & 1.1222 $\pm$ 0.0139 & 1.0765 $\pm$ 0.0008 \\ 
11.1607 & 0.3166 $\pm$ 0.0272 $\pm$ 0.0122 & 0.3408 $\pm$ 0.0293 $\pm$ 0.0132 & 1.1206 $\pm$ 0.0158 & 1.0765 $\pm$ 0.0008 \\ 
11.1677 & 0.3221 $\pm$ 0.0200 $\pm$ 0.0123 & 0.3467 $\pm$ 0.0216 $\pm$ 0.0132 & 1.1247 $\pm$ 0.0134 & 1.0765 $\pm$ 0.0008 \\ 
11.1697 & 0.3569 $\pm$ 0.0196 $\pm$ 0.0135 & 0.3842 $\pm$ 0.0211 $\pm$ 0.0145 & 1.1231 $\pm$ 0.0114 & 1.0765 $\pm$ 0.0008 \\ 
11.1787 & 0.3760 $\pm$ 0.0197 $\pm$ 0.0135 & 0.4047 $\pm$ 0.0212 $\pm$ 0.0146 & 1.1220 $\pm$ 0.0126 & 1.0765 $\pm$ 0.0008 \\ 
11.1812 & 0.3146 $\pm$ 0.0182 $\pm$ 0.0122 & 0.3387 $\pm$ 0.0196 $\pm$ 0.0131 & 1.1214 $\pm$ 0.0121 & 1.0765 $\pm$ 0.0008 \\ 
11.1847 & 0.3454 $\pm$ 0.0196 $\pm$ 0.0134 & 0.3719 $\pm$ 0.0210 $\pm$ 0.0144 & 1.1188 $\pm$ 0.0123 & 1.0765 $\pm$ 0.0008 \\ 
11.1912 & 0.3437 $\pm$ 0.0195 $\pm$ 0.0133 & 0.3700 $\pm$ 0.0210 $\pm$ 0.0144 & 1.1224 $\pm$ 0.0111 & 1.0765 $\pm$ 0.0008 \\ 
11.1942 & 0.3257 $\pm$ 0.0182 $\pm$ 0.0122 & 0.3507 $\pm$ 0.0196 $\pm$ 0.0131 & 1.1258 $\pm$ 0.0115 & 1.0765 $\pm$ 0.0008 \\ 
11.2017 & 0.3687 $\pm$ 0.0204 $\pm$ 0.0134 & 0.3969 $\pm$ 0.0219 $\pm$ 0.0144 & 1.1314 $\pm$ 0.0160 & 1.0765 $\pm$ 0.0008 \\ 
11.2062 & 0.3567 $\pm$ 0.0219 $\pm$ 0.0133 & 0.3840 $\pm$ 0.0236 $\pm$ 0.0144 & 1.1354 $\pm$ 0.0191 & 1.0765 $\pm$ 0.0008 \\ 

  \end{longtable}

\end{center}


\begin{thebibliography}{**}

\bibitem{experiments}
  V.~V.~Ezhela, S.~B.~Lugovsky and O.~V.~Zenin,
  hep-ph/0312114.

\bibitem{Davier}
  M.~Davier, A.~Hoecker, B.~Malaescu and Z.~Zhang,
  Eur.\ Phys.\ J.\ C {\bf 80}, no. 3, 241 (2020).

\bibitem{Teubner} 
  A.~Keshavarzi, D.~Nomura and T.~Teubner,
  Phys.\ Rev.\ D {\bf 101}, no. 1, 014029 (2020).

\bibitem{ref-ask-3-fred} 
  F.~Jegerlehner,
  EPJ Web Conf.\  {\bf 166}, 00022 (2018).

\bibitem{babar_rb}
  B.~Aubert {\it et al.} [BaBar Collaboration],
  Phys.\ Rev.\ Lett.\  {\bf 102}, 012001 (2009).

\bibitem{belle_rb}
  D.~Santel {\it et al.} [Belle Collaboration],
  Phys.\ Rev.\ D {\bf 93}, no. 1, 011101 (2016).

\bibitem{CUSB}
  D.~M.~J.~Lovelock {\it et al.},
  Phys.\ Rev.\ Lett.\  {\bf 54}, 377 (1985).

\bibitem{CLEO}
  D.~Besson {\it et al.} [CLEO Collaboration],
  Phys.\ Rev.\ Lett.\  {\bf 54}, 381 (1985).

\bibitem{rad} E.~A.~Kuraev and V.~S.~Fadin,
Sov.\ J.\ Nucl.\ Phys.\ {\bf 41}, 466 (1985) [Yad.\ Fiz.\ {\bf
41}, 733 (1985).]

\bibitem{Berends}F.~A.~Berends, ``Z Line Shape'',
          CERN {\bf 89-08} (1989), edited by G.~Altarelli,
          R.~Kleiss and C.~Verzegnassi.

\bibitem{ref-ask-2-mont} 
  G.~Montagna, O.~Nicrosini, F.~Piccinini and L.~Trentadue,
  Nucl.\ Phys.\ B {\bf 452}, 161 (1995).
  
\bibitem{Dong:2017tpt} 
  X.~K.~Dong, L.~L.~Wang and C.~Z.~Yuan,
  Chin.\ Phys.\ C {\bf 42}, no. 4, 043002 (2018).
  
\bibitem{Greiner:1994}W.~Greiner, J.~Reinhardt.
Quantum Electrodynamics (3rd Ed.), Springer-Verlag, Berlin, 1994.

\bibitem{Berends_vp} F.~A.~Berends and G.~J.~Komen,
  Phys.\ Lett.\  {\bf 63B}, 432 (1976).

\bibitem{Davier:2002dy}
M.~Davier, S.~Eidelman, A.~Hocker and Z.~Zhang,
Eur.\ Phys.\ J.\ C {\bf 27}, 497 (2003).

\bibitem{PDG}
  M.~Tanabashi {\it et al.} [Particle Data Group],
  Phys.\ Rev.\ D {\bf 98}, no. 3, 030001 (2018).

\bibitem{Bai:1999pk}
J.~Z.~Bai {\it et al.}  [BES Collaboration],
Phys.\ Rev.\ Lett.\  {\bf 84}, 594 (2000).

\bibitem{Bai:2001ct}
J.~Z.~Bai {\it et al.}  [BES Collaboration],
Phys.\ Rev.\ Lett.\  {\bf 88}, 101802 (2002).

\bibitem{Rodrigo:1997zd}
G.~Rodrigo, A.~Pich and A.~Santamaria,
Phys.\ Lett.\ B {\bf 424}, 367 (1998).

\bibitem{ref-KEDR1} 
  V.~V.~Anashin {\it et al.} [KEDR Collaboration],
  Phys.\ Lett.\ B {\bf 788}, 42 (2019).
  
\bibitem{ref-KEDR2} 
  V.~V.~Anashin {\it et al.},
  Phys.\ Lett.\ B {\bf 770}, 174 (2017).

\bibitem{lowess} William S. Cleveland (Wadsworth,
555 Morego Street, Monterey, California 93940), ``The Elements
of Graphing Data''. We use the program from wikipedia
(https://en.wikipedia.org/wiki/Local\_regression) supplied by the
author.

\bibitem{supple} See supplemental material at http://cpc.ihep.ac.cn/article/doi/10.1088/1674-1137/44/8/083001
for the final results shown in the Appendix
and the covariances of the $\RbB$ and $\Rb^{\rm dre}$.

\bibitem{smo_spline}
https://www.mathworks.com/help/curvefit/smoothing-splines.html

\bibitem{radcorr}
  S.~Actis {\it et al.} [Working Group on Radiative Corrections and Monte Carlo Generators for Low Energies],
  Eur.\ Phys.\ J.\ C {\bf 66}, 585 (2010) and references therein.

\bibitem{vacuum} Fred Jegerlehner,
http://www-com.physik.hu-berlin.de/\~\,fjeger/software.html and
the references therein.

\bibitem{ignatov} Fedor Ignatov, talk at the fourth meeting
of the Working Group on Rad. Corrections and MC Generators for Low
Energies, Beijing, October 9-11, 2008.
http://www.lnf.infn.it/wg/sighad/beijing08/Sighadmeeting/sighad08\_vpol.pdf.

\bibitem{HMNT}
  K.~Hagiwara, A.~D.~Martin, D.~Nomura and T.~Teubner,
  Phys.\ Rev.\ D {\bf 69}, 093003 (2004);
  K.~Hagiwara, A.~D.~Martin, D.~Nomura and T.~Teubner,
  Phys.\ Lett.\ B {\bf 649}, 173 (2007).

\bibitem{Y10750} 
  R.~Mizuk {\it et al.} [Belle Collaboration],
  JHEP {\bf 1910}, 220 (2019).

\bibitem{zhuk}
  K.~Zhu, X.~H.~Mo, C.~Z.~Yuan and P.~Wang,
  Int.\ J.\ Mod.\ Phys.\ A {\bf 26}, 4511 (2011).

\bibitem{zhangal} 
  See, for example, B.~Chen, A.~Zhang and J.~He,
  Phys.\ Rev.\ D {\bf 101}, no. 1, 014020 (2020), and references therein.

\bibitem{maiani} 
  A.~Ali, L.~Maiani, A.~Y.~Parkhomenko and W.~Wang,
  Phys.\ Lett.\ B {\bf 802}, 135217 (2020).

\bibitem{zhongxh} 
  Q.~Li, M.~S.~Liu, Q.~F.~Lü, L.~C.~Gui and X.~H.~Zhong,
  Eur.\ Phys.\ J.\ C {\bf 80}, no. 1, 59 (2020).

\bibitem{wangzg} 
  Z.~G.~Wang,
  Chin.\ Phys.\ C {\bf 43}, no. 12, 123102 (2019).

\bibitem{liangwh} 
W.~H.~Liang, N.~Ikeno and E.~Oset,
Phys. Lett. B \textbf{803}, 135340 (2020).

\bibitem{Tornqvist:1984fx}N.~A.~Tornqvist,
Phys.\ Rev.\ Lett.\  {\bf 53}, 878 (1984).

\bibitem{pipihb}
  A.~Abdesselam {\it et al.} [Belle Collaboration],
  Phys.\ Rev.\ Lett.\  {\bf 117}, no. 14, 142001 (2016).

\bibitem{ref-ask-1-kloe} 
  A.~Anastasi {\it et al.} [KLOE-2 Collaboration],
  Phys.\ Lett.\ B {\bf 767}, 485 (2017).

\end{thebibliography}
\end{document}